\documentclass[11pt]{article}
\usepackage{graphicx} 
\usepackage[a4paper, margin = 1.0in]{geometry}
\usepackage{setspace}
\usepackage[section]{placeins}
\usepackage{amsfonts}
\usepackage{amsmath}
\usepackage{bm}
\usepackage{float}
\usepackage[round]{natbib}
\usepackage{authblk}
\usepackage{url}
\usepackage{parskip}
\usepackage{booktabs}
\usepackage{float}
\usepackage{caption}
\usepackage{subcaption}
\floatstyle{plaintop}
\restylefloat{table}
\usepackage{threeparttable}
\usepackage{xcolor}
\usepackage{hyperref}
\usepackage{xr-hyper}
\definecolor{cb-green-sea}  {RGB}{  0, 146, 146}
\definecolor{cb-burgundy}   {RGB}{146,   0,   0}
\definecolor{cb-green-lime} {RGB}{ 36, 255,  36}
\usepackage{amssymb}
\usepackage{pifont}

\makeatletter
\renewcommand{\maketitle}{\bgroup\setlength{\parindent}{0pt}
\begin{flushleft}
  \textbf{\@title}

  \@author
\end{flushleft}\egroup
}
\makeatother
\renewenvironment{abstract}
 {\par\noindent\textbf{\abstractname}\ \ignorespaces}
 {\par\medskip}
 
\title{\Large\textbf{Hilbert space methods for approximating multi-output latent variable Gaussian processes}}
\author[1,3,*]{\small Soham Mukherjee}
\author[2,3]{\small Manfred Claassen}
\author[1]{\small Paul-Christian B\"urkner}
\affil[1]{\footnotesize Department of Statistics, Technische Universit\"at Dortmund, Germany}
\affil[2]{\footnotesize Department of Internal Medicine, Universit\"atsklinikum T\"ubingen, Germany}
\affil[3]{\footnotesize Department of Computer Science, University of T\"ubingen, Germany}

\affil[*]{\footnotesize \textbf{Address for correspondence}: Soham Mukherjee, Department of Statistics, Technische Universit\"at Dortmund, Vogelpothsweg 87, 44227 Dortmund. Email: sohammukherjee1998@gmail.com}
\date{}
\begin{document}

\maketitle 
\begin{abstract}\\
    Gaussian processes are a powerful class of non-linear models, but have limited applicability for larger datasets due to their high computational complexity. In such cases, approximate methods are required, for example, the recently developed class of Hilbert space Gaussian processes. They have been shown to significantly reduce computation time while retaining most of the favorable properties of exact Gaussian processes. However, Hilbert space approximations have so far only been developed for uni-dimensional outputs and manifest (known) inputs. Thus, we generalize Hilbert space methods to multi-output and latent input settings. Through extensive simulations, we show that the developed approximate Gaussian processes are indeed not only faster, but also provide similar or even better uncertainty calibration and accuracy of latent variable estimates compared to exact Gaussian processes. While not necessarily faster than alternative Gaussian process approximations, our new models provide better calibration and estimation accuracy, thus striking an excellent balance between trustworthiness and speed. We additionally illustrate our methods on a real-world case study from single cell biology. 
\end{abstract}

\section{Introduction} \label{sec-intro}
Gaussian processes (GPs) are a popular class of non-linear models well-known for their flexibility in modeling a wide range of data scenarios \citep{Rasmussen1995}. They are also well-known for their steep computational complexity for model fitting, scaling cubically with the sample size. Thus, exact GPs cannot be used to model large datasets within feasible time. These scalability issues are even more pronounced for extensions of standard GP models. Here, we consider a combination of GPs for latent variable estimation \citep{lawrence_gaussian_2003, lawrence_probabilistic_2005} and GPs with multi-dimensional outputs \citep{Cressie1993, teh_semiparametric_2005}. Multi-output GPs are tasked to jointly model several, potentially correlated outcome variables, each with their own GP. In turn, estimating latent variables via GPs not only increases the number of estimable parameters drastically, but also requires multiple (potentially many) outputs to achieve sufficient estimation accuracy. Multi-output GPs for latent variable estimation have important applications, for example, in single-cell biology where estimating latent cell orderings from multi-dimensional RNA gene sequencing data is crucial to understand the dynamics of the biological process \citep{haque_practical_2017}. Yet, when using exact GPs for this purpose, only small datasets can be analyzed, which strongly limits their practical applicability.

Several methods have been proposed as scalable approximations for multi-output GPs \citep{JMLR:v12:alvarez11a, hensman_gaussian_2013} and latent variable GPs \citep{titsias_variational_2009, titsias_bayesian_2010}. These methods primarily depend on reduced rank representations that approximate the GP covariance matrix using inducing points \citep{seeger_fast_2003, snelson_sparse_2005, quinonero2005unifying} followed by approximate model inference via mean-field variational inference (VI) (\cite{titsias_variational_2009}. These GPs scale linearly with sample size and quadratically only with the number of inducing points. Thus, they are computationally efficient as long as the number of inducing points is not too large. However, the lack of support for custom prior specifications of GP hyperparameters across multiple outputs. Although GP parameters can be fixed to a certain value, without varying custom priors, applicability for such models can be limiting for complex data scenarios. Moreover, an in-depth study of the statistical properties of latent variable estimates using these methods are, to the best of our knowledge, yet to be discussed.

An alternative approach is to approximate the covariance function through its spectral decomposition computed from a finite set of representative basis functions. The latter method falls under the category of Hilbert space approximate GPs \citep[HSGPs; ][]{solin_hilbert_2020, riutort-mayol_approx_gps_2022}. By exploiting the spectral representation of a stationary covariance function, the computational complexity of HSGPs scales linearly with both sample size and the number of basis functions. What is more, they come with powerful diagnostics that indicate whether the chosen number of basis functions was sufficient for an accurate GP approximation given the data at hand \citep{riutort-mayol_approx_gps_2022}. HSGPs are typically estimated via Markov-chain Monte Carlo (MCMC), which is slower than VI but produces posterior approximations of much higher quality in terms of accuracy and posterior uncertainty estimation \citep{yao2018yes}. The increased, diagnosable accuracy of HSGPs combined with their efficiency has a lot of potential for latent variable estimation. However, HSGPs have so far only been developed for single outputs and manifest (known) inputs. We develop extensions for the HSGP framework to address both latent variable inputs and multi-dimensional outputs and compare their benefits over exact GPs as well as other GP approximation methods.

\subsection{Overview of Contributions} \label{sec-contrib}

\begin{itemize}
    \item We generalize HSGPs for latent variable inputs and multi-dimensional outputs. 
    \item We validate our new HSGPs against the corresponding exact GPs (where feasible) as well as against variational GPs based on inducing points approximation through a wide range of simulation scenarios.
    \item Our extensive simulation studies investigate the statistical properties of HSGPs in terms of latent variable estimation accuracy, uncertainty calibration, and model convergence.
    \item We illustrate the applicability of our HSGPs on real-world single-cell RNA sequencing data, whose analysis is infeasible with similarly specified exact GPs.
\end{itemize}

\section{Related work} \label{sec-rel-works}
Among the extensions of GPs, previous works discuss exact multi-output GPs \citep{teh_semiparametric_2005} as well as their scalable counterparts \citep{JMLR:v12:alvarez11a, hensman_gaussian_2013} for high-dimensional data. Recent works in multi-output GPs \citep{moreno-munoz_heterogeneous_2018, joukov_fast_2022} also account for output-dimension specific information using different GP hyperparameters for each output dimension. In case of latent variable GPs \citep{lawrence_gaussian_2003, lawrence_probabilistic_2005} current approximations use a combination of inducing points \citep{snelson_sparse_2005, quinonero2005unifying} and mean-field VI \citep{titsias_variational_2009, titsias_bayesian_2010} for scalable solutions. A combination of the above approximation methods for multi-output and latent variable GPs are presented in \cite{lalchand_generalised_2022}. In this paper, we consider HSGPs \citep{solin_hilbert_2020, riutort-mayol_approx_gps_2022} and extend it for multi-dimensional outputs and latent variable inputs. Recently, \cite{mukherjee2025dgplvmderivativegaussianprocess} developed exact multi-output latent variable GPs with derivative information. We only take inspirations from the relevant parts (multi-output and latent inputs) of that work for our model development and experimental design.

In terms of real-world applications, we highlight the field of single-cell biology. There, multi-output structures naturally arise in a plethora of research problems \citep{haque_practical_2017} and latent GPs are common to model cellular ordering \citep{hensman_hierarchical_2013, campbell_bayesian_2015}. Later works involve estimating the cellular ordering \citep{pseudotime_estimation_reid_john_wernisch, campbell_descriptive_2018} along with inference on different cellular pathway dynamics. A direct combination of the above multi-output latent GPs for single-cell RNA sequencing data is presented in \cite{ahmed_grandprix_2019}, which used the combination of inducing points and VI methods for estimating cellular ordering. As a real-world case study, we showcase our models for a single-cell data of the cell cycle \citep{mahdessian_spatiotemporal_2021}.

\section{Methods} \label{sec-methods}
We initially discuss the structure of exact GPs and HSGPs for univariate output $y$ and covariate (known) with paired samples $(y_i, x_i)$ for $i = {1, 2, ..., N}$ where $N$ is the sample size. Then, we generalize the Hilbert space framework for multi-dimensional outputs and latent inputs before discussing methods of inference and alternate GP approximation methods.

\subsection{Gaussian processes} \label{sec-gp}
A Gaussian process (GP) is a stochastic process specified by a mean function $\mu = \mu(x)$ and a covariance function $k = k(x, x')$ where $x, x'$ are two arbitrary input values. We write $f \sim \mathcal{GP}(\mu, k)$ to indicate that $f = f(x)$ is distributed as a GP such that any finite subset of $f$ jointly follows a multivariate Gaussian distribution. For a univariate output $y$, a standard Gaussian process regression is given by
\begin{equation} \label{eq-gpr}
    y_i = f(x_i) + \varepsilon_i
\end{equation}
where $\varepsilon_i$ is the $i^{th}$ sample of the independent additive noise $\varepsilon$ with error variance $\sigma^2$ such that $\varepsilon \sim \mathcal{N}(0, \sigma^2)$. The entire model can thus be written as 
 \begin{equation} \label{eqn-gpr-alt}
     y_i \mid f \sim \mathcal{N}(f(x_i), \sigma^2).
 \end{equation}
For $i \neq j$, $\text{Cov}(y_i, y_j) = k(x_i, x_j)$ such that $i,j = {1,2,...,N}$. When $i = j$, $\text{Cov}(y_i, y_j) = \text{Var}(y_i) = k(x_i,x_j) + \sigma^2$. In terms of $k$, the Matern class of stationary covariance functions is a highly common choice \citep{Rasmussen1995, gp_rasmussen_williams_2006}. Among them, we specifically consider Squared Exponential (SE), Matern 3/2 and 5/2, which are defined as:
\begin{equation} \label{eqn-covfns}
    \begin{aligned}
        k_{\rm se} &= \alpha^2 \exp \left(-\frac{r^2}{2\rho^2}\right), & \\
        k_{\rm m32} &= \alpha^2 \left(1 + \frac{\sqrt{3r^2}}{\rho} \right) \exp\left(-\frac{\sqrt{3r^2}}{\rho}\right), & \\
        k_{\rm m52} &= \alpha^2 \left(1 + \frac{\sqrt{5r^2}}{\rho} + \frac{5r^2}{3\rho^2} \right) \exp\left(-\frac{\sqrt{5r^2}}{\rho}\right),
    \end{aligned}
\end{equation}
where $r = | x - x' |$ is the distance between two arbitrary inputs, $\alpha > 0$ is the marginal standard deviation (SD) and $\rho > 0$ is the length scale parameter. The above covariance functions vary in terms of their implied functional smoothness, with the SE covariance function being the most and the Matern 3/2 being the least smooth. We use a constant mean function $\mu$ (as an estimated parameter independent of inputs) for all of our exact GPs and HSGPs discussed in this paper.

\subsection{Hilbert space approximations} \label{sec-hilbert-approx} 

In the HSGP framework \citep{solin_hilbert_2020, riutort-mayol_approx_gps_2022}, stationary covariance functions are approximated using their spectral densities in the frequency domain. Briefly, Bochner's theorem \citep{Gihman2004} states that a covariance function of a stationary process can be expressed as the Fourier transform of a positive finite measure. If that measure has a density, then we call it a spectral density of the covariance function. Further, using Wiener-Khintchine theorem \citep{AnalTimeSeriesChatfield}, we can show that a covariance function and its spectral density are Fourier duals. Combining the results from both these theorems, any arbitrary function satisfying the criteria of being a covariance function (symmetric and positive-definite) can be expressed in terms of its spectral density. The spectral densities corresponding to the covariance functions from Eq.\eqref{eqn-covfns} are given by
\begin{equation} \label{eqn-specdens}
    \begin{aligned}
        S_{\rm se}(\omega) &= \alpha^2\rho(\sqrt{2\pi})\exp\left(-\frac{1}{2}\rho^2\omega^T\omega\right), &\\
        S_{\rm m32}(\omega) &= \alpha^2\left(\frac{2\Gamma(2)3^{3/2}}{\rho^3/2}\right)\exp\left(\frac{3}{\rho^2}+\omega^T\omega\right)^{-2}, &\\
        S_{\rm m52}(\omega) &= \alpha^2\left(\frac{2\Gamma(3)5^{5/2}}{3\rho^5/4}\right)\exp\left(\frac{5}{\rho^2}+\omega^T\omega\right)^{-3},
    \end{aligned}
\end{equation}
where $\omega\in \mathbb{R}^N$ is the input in the frequency domain, and $\rho > 0$ and $\alpha > 0$ are the covariance function hyperparameters. The spectral densities can be derived for the general Matern class of covariance functions \citep{riutort-mayol_approx_gps_2022}, but for this paper, we specifically focus on the above special cases. 

We follow the procedure of \cite{riutort-mayol_approx_gps_2022} and start by defining the compact cover $\Omega \in [-L, L] \subset \mathbb{R}$ containing each element $x$ from the input space for any positive real number $L$. Following this, we can write any stationary covariance function with $x, x' \in \Omega$ as
\begin{equation} \label{eqn-covfn-specdens}
    k_{\theta}(x, x') = \sum_{j = 1}^{\infty}S_{\theta}(\sqrt{\lambda_j})\phi_j(x)\phi_j(x'),
\end{equation}
where $S_{\theta}$ is the spectral density of the covariance function $k_{\theta}$ with $\theta$ being the set of hyperparameters. The sets of eigenvalues $\{\lambda_j\}_{j=1}^{\infty}$ and eigenfunctions $\{\phi_j\}_{j = 1}^{\infty}$ of the Laplacian operator in the domain $\Omega$ satisfy the following problem in $\Omega$ under Dirichlet boundary conditions \citep{solin_hilbert_2020}:
\begin{equation} \label{eqn-bound-conds}
    \begin{aligned}
        -\nabla^2\phi_j(x) &= \lambda_j \phi_j(x), \quad x \in \Omega &\\ 
        \phi_j(x) &= 0, \quad x \notin \Omega 
    \end{aligned}
\end{equation}
where the eigenvalues $\lambda_j > 0$ are real positive due to the Laplacian being a positive-definite Hermitian operator. Here, the $\phi_j$ are sinusoidal functions and the solution to the eigenvalue problem in Eq.\eqref{eqn-bound-conds} are independent of the choice of covariance functions. They are given by
\begin{equation} \label{eqn-eigen-solns}
    \begin{aligned}
        \lambda_j &= \left(\frac{j\pi}{2L}\right)^2, &\\
        \phi_j(x) &= \sqrt{\frac{1}{L}} \sin\left(\sqrt{\lambda_j}(x + L)\right).
    \end{aligned}
\end{equation}
We approximate the covariance function using the linear combination of its basis functions. Considering the first $M$ number of basis terms, we obtain from Eq.\eqref{eqn-covfn-specdens}, 
\begin{equation} \label{eqn-covfn-approx}
    k_{\theta}(x, x') \approx \sum_{j = 1}^{M}S_{\theta}(\sqrt{\lambda_j})\phi_j(x)\phi_j(x').
\end{equation}
The full covariance matrix $K$ for inputs $x_i$, $i \in {1, ..., N} $ is thus approximated by its finite basis function as
\begin{equation} \label{eqn-covmat-eigendecomp}
    K \approx \Phi\Delta\Phi^T
\end{equation}
where $\Delta = \text{diag}\{S_{\theta}(\sqrt{\lambda_1}), \ldots,S_{\theta}(\sqrt{\lambda_M})\}$ is the diagonal matrix containing spectral densities with eigenvalues and $\Phi \in \mathbb{R}^{N\times M}$ is the matrix of eigenfunctions
\begin{equation} \label{eqn-eigenmatrix}
    \Phi = \Phi(\mathbf{x}) = \begin{bmatrix}
    \phi_1(x_1) & \dots & \phi_M(x_1) \\
    \vdots & \ddots & \vdots \\
    \phi_1(x_N) & \dots & \phi_M(x_N)
    \end{bmatrix}.
\end{equation}
for a vector of input points $\mathbf{x} = (x_1, \ldots, x_N)$.
We can thus re-write the GP functions from Section \ref{sec-gp} as 
\begin{equation} \label{eqn-gp-eigen}
    f \sim \mathcal{N}(\mu, \Phi\Delta\Phi^T).
\end{equation}
This is equivalent to the linear representation 
\begin{equation} \label{eqn-gp-approx}
    f(\mathbf{x}) \approx \mu + \sum^{M}_{j = 1} \left(S_{\theta}(\sqrt{\lambda_j})\right)^{1/2} \Phi(\mathbf{x}) \beta_j 
\end{equation}
where $\beta_j \sim \mathcal{N}(0, 1)$ is standard normal. The eigenvalues $\lambda_j$ are monotonically increasing in $j$ and $S_{\theta}$ rapidly decreases to zero for a bounded covariance functions. As a result, usually only a small number of basis functions $M$ are required for an adequate approximation \citep{gp_rasmussen_williams_2006, riutort-mayol_approx_gps_2022}.

\subsection{Extending HSGPs} \label{sec-mult-latent}
We extend the HSGP framework to model multi-dimensional outputs and latent variable inputs. We follow a similar framework pertaining to the multi-output latent variable models considered in \cite{mukherjee2025dgplvmderivativegaussianprocess} (but do not consider the derivative methods here). 

\subsubsection{Multi-output HSGPs} \label{sec-multiout}
We specify multi-output GPs with response variables $(y_1,\dots,y_D)$ over $D>1$ output dimensions \citep{gp_rasmussen_williams_2006}, using $y_{di}$ to denote the response for dimension $d$ and observation $i$. As the usual approach, we first set up $D$ independent, univariate Gaussian processes $f_d$ each with their own set of hyperparameters, say $\theta_d$ \citep{teh_semiparametric_2005, mukherjee2025dgplvmderivativegaussianprocess}. Thus, extending HSGPs to multi-output GPs, we first modify Eq.\eqref{eqn-gp-approx} such that
\begin{equation} \label{eqn-gp-approx-multi}
    f_d(\mathbf{x}) \approx \mu_d + \sum^{M}_{j = 1} \left(S_{\theta_d}(\sqrt{\lambda_j})\right)^{1/2} \Phi(\mathbf{x}) \beta_{jd}
\end{equation} 
where $\beta_{jd} \sim \mathcal{N}(0, 1)$. The univariate GPs are then related to one another by folding them with a ($D$-dimensional) across-dimension correlation matrix $C$ \citep{teh_semiparametric_2005, bonilla_multi-task_2007}. Specifically, for each observation $i$, we obtain a vector of across-dimension correlated GP values as
\begin{equation}
\label{multi-gps}
\left(\begin{array}{c} f^*_1(x_i) \\ \dots \\ f^*_D(x_i) \end{array}\right) 
= A \times \left(\begin{array}{c} f_1(x_i) \\ \dots \\ f_D(x_i) \end{array}\right),
\end{equation}
where $A$ is the Cholesky factor of $C$ such that $C = AA^T$ with $A$ being lower-triangular. The Cholesky factor $A$ is matrix multiplied ($\times$) to the vector of across-dimension GP values. This way, multi-output GPs combine two dependency structures, one within dimensions (and across observations) as expressed by the univariate GPs through corresponding covariance functions and one across output dimensions (but within observations) as expressed by $C$ (or $A$). We specify $C$ through a Cholesky prior $C \sim\text{LKJ}(\eta = 1$), which is uniform across all valid correlation matrices of a given dimension \citep{LKJcholesky2009}. For efficiency reasons, we equivalently estimate $A$ instead of $C$ via MCMC along with all other model parameters. More details on model setup are discussed later in \ref{sec-model-spec}.  Adding independent Gaussian noise to our derivative multi-output GP model, we extend Eq.\eqref{eqn-gpr-alt} for all $d$ and $i$:
\begin{equation} \label{multi-GPs}
y_{di} \mid f^*_d \sim \mathcal{N}(f^*_d(x_i), \sigma_d^2) 
\end{equation}
An alternative way of specifying $C$ is through a set of input-dependent covariance functions \citep{Gneiting2010crosscovariance, Bourotte2016nonseparablecrosscovariance}. In our study, we however only consider an input-independent $C$, as modeling the across-dimension correlations is not the main focus of this paper. In Section \ref{sec-discuss}, we provide a more detailed discussion on this topic.

\subsubsection{Latent variable HSGPs}\label{sec-latent}
Within latent-variable GPs, the inputs $x$ are considered as unobserved and are treated similar to other estimable parameters. To that end, we specify an observed quantity $\tilde{x}$ that guides the estimation of the latent inputs $x$. From a Bayesian perspective, $\tilde{x}$ acts as a prior-like object for the latent $x$ to be then further refined by the GPs learning from $y$. Specifically, we consider the scenario where $\tilde{x}$ is a noisy measurement of $x$. If we assume that the measurements $\tilde{x}$ are Gaussian with known measurement SD $s$, we can write for each observation $i$:
\begin{equation} \label{eqn-latent prior}
    \tilde{x}_i \sim \mathcal{N}(x_i, s^2).
\end{equation}
The vector of latent inputs $\mathbf{x} = (x_i, \ldots, x_N)$ is then passed to the approximation step in Eq.\eqref{eqn-gp-approx} (and subsequently Eq.\eqref{eqn-gp-approx-multi} for the multi-output case). We call the resulting model (multi-output) latent variable HSGPs.

Latent variable GPs, exact or approximate, are more difficult to fit than their manifest counterparts. The primary reasons are the substantial increase in the number of estimable parameters as well as identification issues arising due to both $x$ and $\rho$ now being treated as unknown. We alleviate this identifiability issue using MCMC methods discussed in Section \ref{sec-mcmc}. We further present a detailed simulation study investigating the challenges of latent variable GPs in Section \ref{sec-sim-study}.  

\subsection{Bayesian inference} \label{sec-mcmc}
We fit HSGPs using full Bayesian inference via MCMC sampling. Our primary parameters of interest are the latent inputs $x$ and GP hyperparameters $\theta$ given data (outputs) $y$. We re-purpose $\theta$ to include not only the covariance function hyperparameters $\rho$ and $\alpha$ as specified earlier but also the error SD $\sigma$. For a specific output dimension $d$, we assume independent prior distributions on $\theta_d$ as 
\begin{equation}\label{eqn-hyperprior}
    \theta_d \sim p(\theta_d) = p(\rho_d) \, p(\alpha_d) \, p(\sigma_d).
\end{equation}
The joint probability density factorizes as
\begin{equation}\label{eqn-jt prob model}
    p(y, x, \theta \mid \tilde{x}) = p(x \mid \tilde{x}) \prod_d^D p(y_d \mid x, \theta_d) \, \, p(\theta_d).
\end{equation}
where $p(y_d \mid x, \theta_d)$ denotes the GP-based likelihood for a single output dimension and $p(x \mid \tilde{x})$ denotes the prior for the latent $x$ implied by the measurement model in Eq.\eqref{eqn-latent prior}. The details of prior specifications used in our experiments are further discussed in Section \ref{sec-sim-study}. Using Bayes' rule, we obtain the joint posterior over $x$ and $\theta$ as
\begin{equation}\label{eqn-post x}
    p(x, \theta \mid y, \tilde{x}) = 
    \frac{p(y, x, \theta \mid \tilde{x})}
    {\int\int p(y, x, \theta \mid \tilde{x}) \, dx \, d\theta}.
\end{equation}
The inference framework specified above is same for both exact and HSGPs. Posterior samples of $x$ and $\theta$ for all output dimensions are obtained via MCMC sampling, specifically adaptive Hamiltonian Monte Carlo \citep{Neal2011HMC, HoffmanM2014NUTS}. We implemented the exact GP and the HSGP models in Stan using the RStan interface \citep{Stan_guide_2024}.

\subsection{Approximate latent variable GPs using VI}
Approximating latent variable GPs using VI methods (VIGPs) usually include a combination of approximating both the covariance function and the posterior. Approximating the covariance function is carried out using inducing points to obtain a reduced rank representation of the Gram matrix \citep{gp_rasmussen_williams_2006}. There are many inducing point methods \citep{quinonero2005unifying}, however, based on the suggestions of \cite{VIGPsWilk2016}, we specifically consider the implementation discussed in \cite{titsias_variational_2009}. In this framework, the estimation of latent variables $x$ and GP hyperparameters $\theta$ after marginalizing over $f$ is carried out via the marginal likelihood
\begin{equation}\label{gplvm-marginal-exact}
    p(y \mid x,\theta) = \mathcal{N}(y \mid \mu, K_{N \times N} + \sigma^2 I),
\end{equation}
where $K_{N\times N}$ is the full rank GP covariance matrix for a sample size $N$. This full rank covariance matrix is then replaced by $Q_{N\times N} = K_{N\times M}K^{-1}_{M\times M}K_{M\times N}$ where $M$ is the number of inducing points. This reduces the computational complexity of inverting a full rank matrix to only computing its Nystr\"om approximation. While fundamentally different, the basis points in our method explained in Section \ref{sec-hilbert-approx} and inducing points here serve a similar purpose. For that reason, we choose to denote both the number of basis points (in HSGPs) and inducing points (in VIGPs) by $M$ for simplicity. Effectively, the marginal likelihood is then reformulated as
\begin{equation}\label{gplvm-marginal-approx}
    p(y \mid x,\theta^*) = \mathcal{N}(y \mid \mu, Q_{N \times N} + \sigma^2 I).
\end{equation}
Following the specifications discussed in \cite{titsias_bayesian_2010}, the joint probability distribution for the $d^{th}$ output dimension is given by
\begin{equation}\label{gplvm-jt-dist}
    p(y_d, f_d, M_d) = p(y_d \mid f_d) \, p(f_d \mid M_d) \, p(M_d),
\end{equation}
where $M_d$ is the dimension-specific set of inducing points. The inducing points are designated as variational parameters and optimized through variational inference \cite{titsias_variational_2009, titsias_bayesian_2010}. The posterior $p(f_d, M_d \mid y_d) = p(f_d \mid M_d, y_d) \, p(M_d \mid y_d)$ is subsequently approximated by a sparse variational distribution of the form
\begin{equation}\label{gplvm-approx-vi}
    q_p(f_d, M_d) = p(f_d \mid M_d) \, q_M(M_d),
\end{equation}
where $q_M()$ is the variational distribution over the inducing points $M_d$. Model inference is carried out by the usual method of computing the variational lower bound (see \cite{titsias_bayesian_2010} for further details). 

While the latent variable GPs were earlier designed for classification tasks (\cite{lawrence_gaussian_2003}, \cite{lawrence_probabilistic_2005}), VIGPs have later been used for GP regressions as well, thus making it an ideal candidate for reference and comparison to our proposed HSGPs. In this paper, we follow the implementations for VIGPs in \cite{titsias_bayesian_2010, ahmed_grandprix_2019} using the GP-LVM module of Pyro \citep{pyro2019} where the user only needs to set the number of inducing points and the internal working takes care of their optimization for sparse latent input estimation.

\section{Simulation Study} \label{sec-sim-study}
For real-world data, we lack the ground truth values of latent variables. In contrast, in a simulated setting, we have full control over the data generating process including the true latent values. Thus, it is crucial to investigate the statistical properties of latent variable models using simulated data. To this end, we investigate exact GPs, HSGPs and VIGPs under various simulation scenarios exploring their behavior in terms of model convergence, model calibration, and estimation accuracy. The data generating conditions are inspired by \cite{mukherjee2025dgplvmderivativegaussianprocess}.

\subsection{Data generating scenarios} \label{sec-sim-data}
We consider six different simulation scenarios. These scenarios are broadly categorized into data generated from a GP and data generated from a periodic function (non-GP data scenario). Under the former category, we consider three data generating processes based on multi-output GPs with SE, Matern 3/2, and Matern 5/2 covariance functions, respectively. We sample length-scale $\rho \sim \text{Normal}^+(1, 0.05^2)$, GP marginal SD $\alpha \sim \text{Normal}^+(3, 0.25^2)$, and error SD $\sigma \sim \text{Normal}^+(1, 0.25^2)$. 
Additionally, we consider a fourth GP data scenario, where we allow the length-scale of the SE covariance function to be highly variable across each output dimensions, $\rho \sim \text{Normal}^+(1, 0.25^2)$, thus increasing the challenges in model fitting. Each of the hyperparameters are sampled $D$ times, one value per output dimension. Under the GP data scenarios, the fitted models align with the true data generating process (or are approximations to them). However, simulating from GPs results in a lot of variation in the true functional forms across output dimensions and simulation trials. 

Compared to that, the periodic data generating process have more consistency since we ensure a similar level of non-linearity across each outputs. Concretely, we use the following periodic process:
\begin{equation} \label{eqn-per-sim}
    \begin{aligned}
        f_{id} &= \alpha_d \sin \left(\frac{x_i}{\rho_d}\right), \\[3pt]
        y_{id} &\sim \mathcal{N}(f_{id}, \sigma_d^2).
    \end{aligned}
\end{equation}
In our simulations, we consider two version of this process, one with low oscillations (high true $\rho_d$ values) and one with higher oscillations (low true $\rho_d$ values). These two periodic data scenarios mimic the challenges of modeling starkly different functional smoothness with the higher oscillations being the more challenging one. The hyperparameters of the periodic process closely resemble that of the GP data scenarios and are thus named the same for simplicity. The sampling distributions are also similarly specified as mentioned previously. For the periodic data with higher oscillation, we specify $\rho \sim \text{Normal}^+(0.5, 0.05^2)$ as compared to the lower oscillation case where $\rho \sim \text{Normal}^+(1, 0.25^2)$. 

In all scenarios, we sample the between-dimension (output) correlation matrix $C \sim\text{LKJ}(\eta = 1$) \citep{LKJcholesky2009}. This implies $C$ to be uniformly distributed within the set of all valid correlation matrices of dimension $D$. For the sample size $N$, we consider three cases $N = 20, 50 \text{ } \text{and} \text{ } 200$. Due to the computational complexity of exact GPs, we only consider them for $N = 20$ case. Further, a special case for the GP data scenarios is also demonstrated for the HSGPs and VIGP with $N = 1000$ in Section \ref{sec-simspl-case}. Unlike the smaller $N$ cases, here the fitted models aren't aligned with the true data generating process, that is, we specify vague priors for our HSGP hyperparameters for $N = 1000$ case. We generate the ground truth latent inputs as $x_i \sim \text{Uniform}(0, 10)$, where $i = 1, \ldots, N$. Further, we assume a prior measurement SD of the noisy $\tilde{x}$ as $s = 0.3$ (see Section \ref{sec-mult-latent}).  For the number of output dimensions $D$, we consider the three cases $D = 5, 10 \text{ } \text{and} \text{ } 20$. We perform 50 trials for each of the choices of $N$ and $D$ per simulation scenario. 

\subsection{Model specifications} \label{sec-model-spec}
We fit exact GPs, HSGPs, and VIGPs for all simulation scenarios, with the exception of omitting exact GPs for sample sizes higher than $N=20$ due to the scalability issues discussed earlier (see Section \ref{sec-intro}). For the HSGPs, we select the boundary conditions $L$ such that it contains the input space completely. This is obtained by multiplying a scalar adjustment $c = 1.25$ \citep{riutort-mayol_approx_gps_2022} to the range of the input prior $\tilde{x}$. Thus, we ensure that we have a compact cover $[-L, L]$ that prevents input values $x$ to be near the boundaries. By doing so, we ensure stable MCMC model convergence for the latent variable inputs (see Section \ref{sec-model-conv}). We select the minimum number of basis functions $M_{\rm min}$ loosely based on the suggestions of \cite{riutort-mayol_approx_gps_2022}. The choices for $M_{\rm min}$ depends on the choice of covariance function to accurately approximate the degree of non-linearity. Thus, for the SE covariance function,
\begin{equation}\label{eqn-m-se}
    M_{\rm min} = 1.75 \frac{cS}{\mu_{\rho}},
\end{equation}
for the Matern 3/2,
\begin{equation}\label{eqn-m-m32}
    M_{\rm min} = 3.42 \frac{cS}{\mu_{\rho}},
\end{equation}
and for the Matern 5/2,
\begin{equation}\label{eqn-m-m52}
    M_{\rm min} = 2.65 \frac{cS}{\mu_{\rho}}.
\end{equation}
In the above equations, $c$ is the scaler adjustment to the boundary conditions, $S$ is the range of inputs and $\mu_{\rho}$ is the mean of the length-scale prior. The SE covariance function requires much lower $M_{\rm min}$ compared to, say, the Matern 3/2 which is on the other end of the smoothness spectrum among the Matern class. 

Prior specifications for the exact GPs and HSGPs are kept same for all model parameters. Our prior choices are aligned with the data generating process since that is required for testing model calibration (see Section \ref{sec-model-calib}). Thus, we specify priors for length-scale $\rho \sim \text{Normal}^+(1, 0.05^2)$, GP marginal SD $\alpha \sim \text{Normal}^+(3, 0.25^2)$ and error SD $\sigma \sim \text{Normal}^+(1, 0.25^2)$. For the periodic data with higher oscillation, we specify a lower prior mean of $\rho \sim \text{Normal}^+(0.5, 0.05^2)$. Lastly, in case of the SE simulation scenario with highly varying length-scale, we specify a prior of $\rho \sim \text{Normal}^+(1, 0.25^2)$. The number of estimated primary parameters of interest for each HSGPs are provided in Table \ref{tab:params}.

\begin{table}[!ht]
     \centering
     \caption{Primary parameters of interest}
     \begin{tabular}{c c c}
     \toprule
      Name & Notation & Number of parameters \\
  \midrule
 Latent inputs & $x$ & $N$ \\ 
 Length scale & $\rho$ & $D$ \\
 GP marginal SD & $\alpha$ & $D$ \\
 Error SD & $\sigma$ & $D$ \\
  \bottomrule
     \end{tabular}
     \begin{tablenotes}
        \item \small \textit{Note: $N$ stands for number of observations or sample size. $D$ denotes the number of output dimensions.}
     \end{tablenotes}
     \label{tab:params}
 \end{table}
 
Following the suggestions of the number of inducing points discussed in \cite{titsias_variational_2009}, we use $M=10$ inducing points for the VIGPs. In this framework, a small constant is added to the diagonal terms of the matrix. This helps maintain numerical stability while computing the GP covariance matrix. In case of VIGPs, we tested different values of diagonal constants starting from $\delta = 10^{-12}$ (usually suggested default). Based on our tests, for 10 inducing variables (points), the VIGPs need at least $\delta = 10^{-3}$ for a numerically positive definite GP covariance matrices and reliably finish an entire simulation scenario. Higher number of inducing points would require an even larger $\delta$. Since $\delta = 10^{-3}$ is already quite large, we keep $M=10$ to not further increase the bias caused by large values of $\delta$. As in the other methods, we set the latent prior measurement SD to $s = 0.3$. As for the GP hyperparameters, since the VIGP implementation in Pyro does not support custom priors we keep them at their uniform defaults for the simulation scenarios with $N = 20, 50 \text{ } \text{and} \text{ } 200$. For the $N = 1000$ (see Section \ref{sec-simspl-case}), we introduce a VIGP case where GP hyperparameters are fixed to the data generating conditions, essentially creating a favorable condition for VIGPs.

Both exact GPs and HSGPs are implemented in Stan \citep{Stan_guide_2024} and fitted with a single MCMC chain of 2000 iterations of which the first 1000 are discarded as warm-up. As shown in \cite{mukherjee2025dgplvmderivativegaussianprocess}, model convergence is similar for multiple chains when fitting latent variable GPs. Thus, we run only a single chain per model to reduce overall computation times. The simulation studies are conducted on 50 vCPUs (Intel(R) Xeon(R) Gold 6230R CPU @ 2.10 GHz) with 720 GB work memory. The average runtime per dataset for the HSGPs with $N = 200$ and $D = 20$, the most computationally expensive case, was approximately 28 minutes for the SE model, 42 minutes for the Matern 3/2 model, and 34 minutes for the Matern 5/2 model. In the same scenarios, the VIGPs took approximately 2 minutes regardless of the covariance function. The exact GPs (just for $N=20$ and $D = 20$) took 48 minutes for the SE model, 1.6 hours for the Matern 3/2 and 2.4 hours for the Matern 5/2 model. The HSGPs were significantly faster than the exact GPs but slower than VIGPs. The slower speed of HSGPs compared to VIGPs is counter-balanced by highly consistent model calibrations, superior estimation accuracy for latent variables, and overall much more stable model performance, as shown below.

\subsection{Model convergence} \label{sec-model-conv}
We investigate MCMC convergence of our fitted exact GPs and HSGPs for all of the six simulation study scenarios discussed before. We use standard MCMC sampling diagnostics including state-of-the-art versions of the scale reduction factor $\widehat{R}$, bulk effective sample size (Bulk-ESS) and tail effective sample size (Tail-ESS)  \citep{RankNorm_Vehtari_etal}. A combined check of these measures provide a comprehensive picture of the parameter-specific model convergence. In general, $\widehat{R}$ should be very close to 1 and should ideally not exceed 1.01 \citep{RankNorm_Vehtari_etal}. We additionally consider a more relaxed threshold of 1.1 in our simulation studies, since the ground truth is available as another layer of evaluation. Bulk-ESS indicates the reliability of measures of central tendency such as the posterior mean or median. Tail-ESS indicates the reliability of the 5\% and 95\% quantile estimates, which are then used to construct credible intervals. Both Bulk-ESS and Tail-ESS should have values greater than 100 times the number of MCMC chains (higher is better). We compute all the convergence measures with the posterior package \citep{posterior2023}. Inference of the VIGPs is done via numerical optimization, which we considered to have converged if no convergence warnings were issued by the optimizer.

\begin{figure}[!ht]
    \centering
    \includegraphics[width = \linewidth]{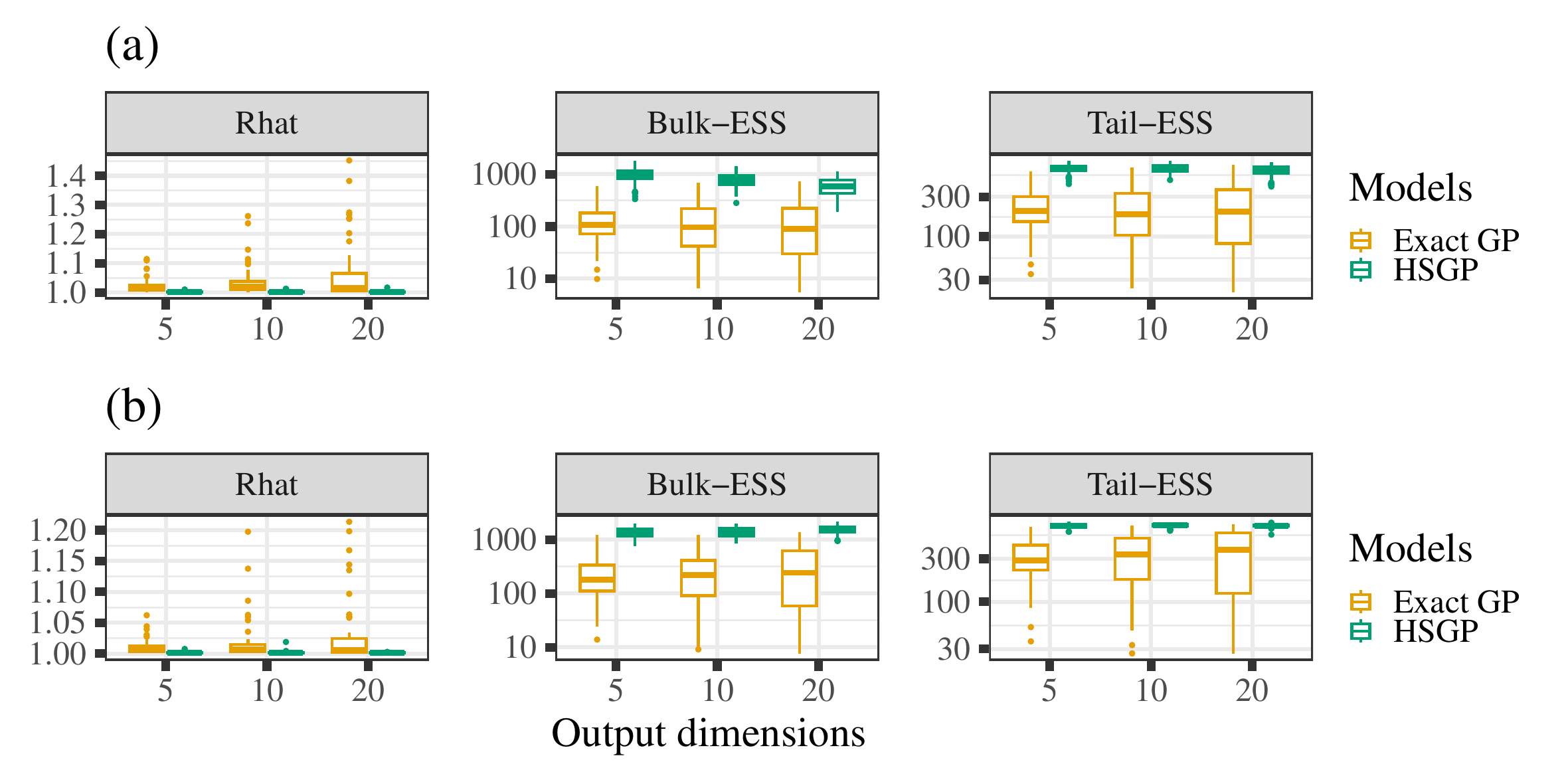}
    \caption{\textit{Squared exponential scenario: Convergence check for (a) latent inputs and (b) GP hyperparameters of the exact GPs (for $N = 20$) and HSGPs (combined for $N = 20, 50 \text{ } \text{and} \text{ } 200$ cases). The y-axes for Bulk and Tail ESS plots are log10 transformed.}}
    \label{fig:se-exact-hsgp-valid}
\end{figure}

In Figure \ref{fig:se-exact-hsgp-valid}, we show $\widehat{R}$, Bulk-ESS, and Tail-ESS for the latent $x$ and GP hyperparameters of both exact GPs and HSGPs with SE covariance function. While the exact GPs reach and exceed the relaxed $\widehat{R}$ threshold of 1.1 for some simulated datasets, the HSGPs consistently satisfy the much stricter 1.01 threshold of model convergence. HSGPs subsequently also have much higher Bulk and Tail-ESS. Overall, based on the diagnostics, HSGPs show much more consistent and stable convergence as compared to exact GPs. We provide convergence diagnostic figures for the other simulation scenarios in the supplementary materials Section C (Figures S3-S7) where we see similar results. We additionally provide further analysis for MCMC convergence of latent inputs and GP hyperparameters compared between exact GPs and HSGPs in supplementary materials Section D (Figures S8-S11).

\subsection{Testing model calibration} \label{sec-model-calib}
We use simulation based calibration (SBC) \citep{modrakSBCcheck2023, taltsValidaBayesInf2020} to test model calibration for estimating latent inputs $x$. Through SBC, we check if the model is able to produce posterior distributions that are consistent with the data generating process, thus correctly accounting for the implied uncertainty.
\begin{figure}[!ht]
    \centering
    \includegraphics[width = \linewidth]{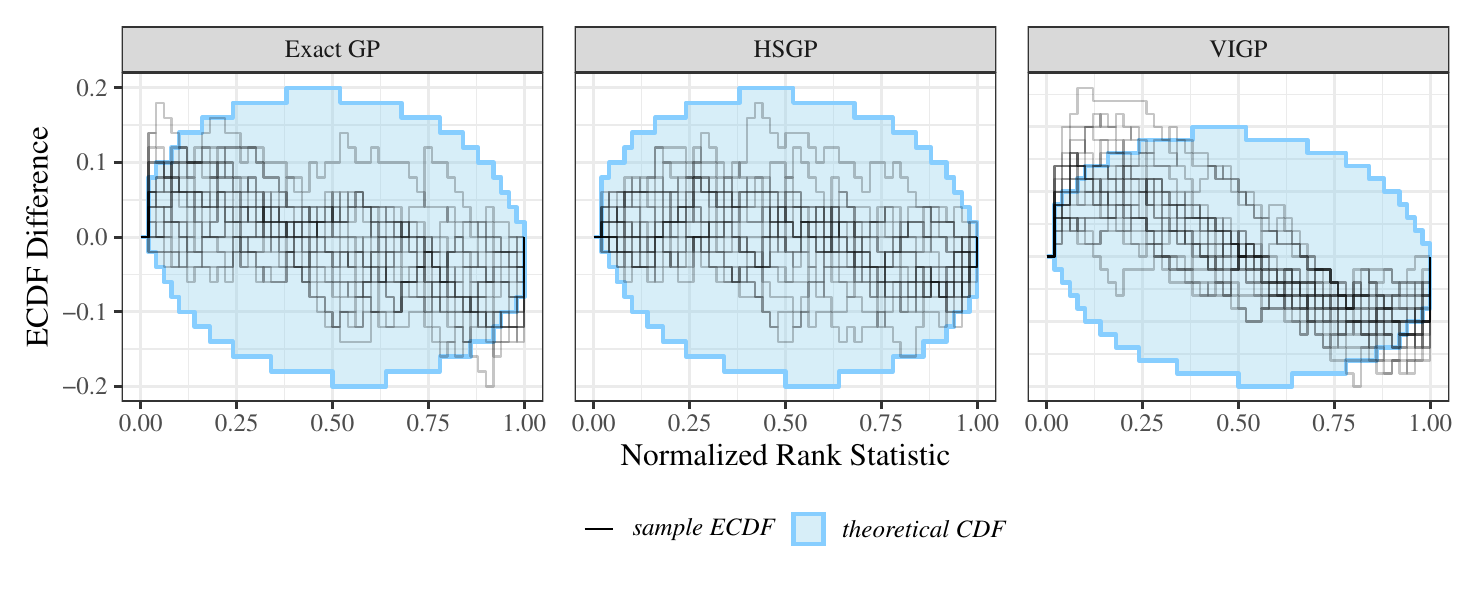}
    \caption{\textit{Squared exponential scenario: ECDF-difference calibration plots of the latent $x$ estimated by exact GP, HSGP, and VIGP. Only the HSGP is consistently well calibrated.}}
    \label{fig:se-ecdf-n20-d20}
\end{figure}
To that end, briefly, SBC aims to test the goodness of posterior approximations by exploiting self-consistency properties of Bayesian models. This results in a rank statistic that is uniformly distributed under the assumption of a well calibrated model \citep{modrakSBCcheck2023}, which can subsequently be checked both graphically and numerically. As a graphical test, we plot the empirical cumulative distribution function (ECDF) of the ranks along with their 95\% confidence regions under the assumption of uniformity \citep{sailynojaGraphUniftest2022}. ECDFs lying outside of their confidence region indicate miscalibrations. These graphical tests also yield a test statistic \citep[the log $\gamma$ score; see][]{sailynojaGraphUniftest2022}, which we further analyze numerically. More details on SBC is provided in supplementary materials Section A. 

\begin{figure}[!ht]
    \centering
    \includegraphics[width = \linewidth]{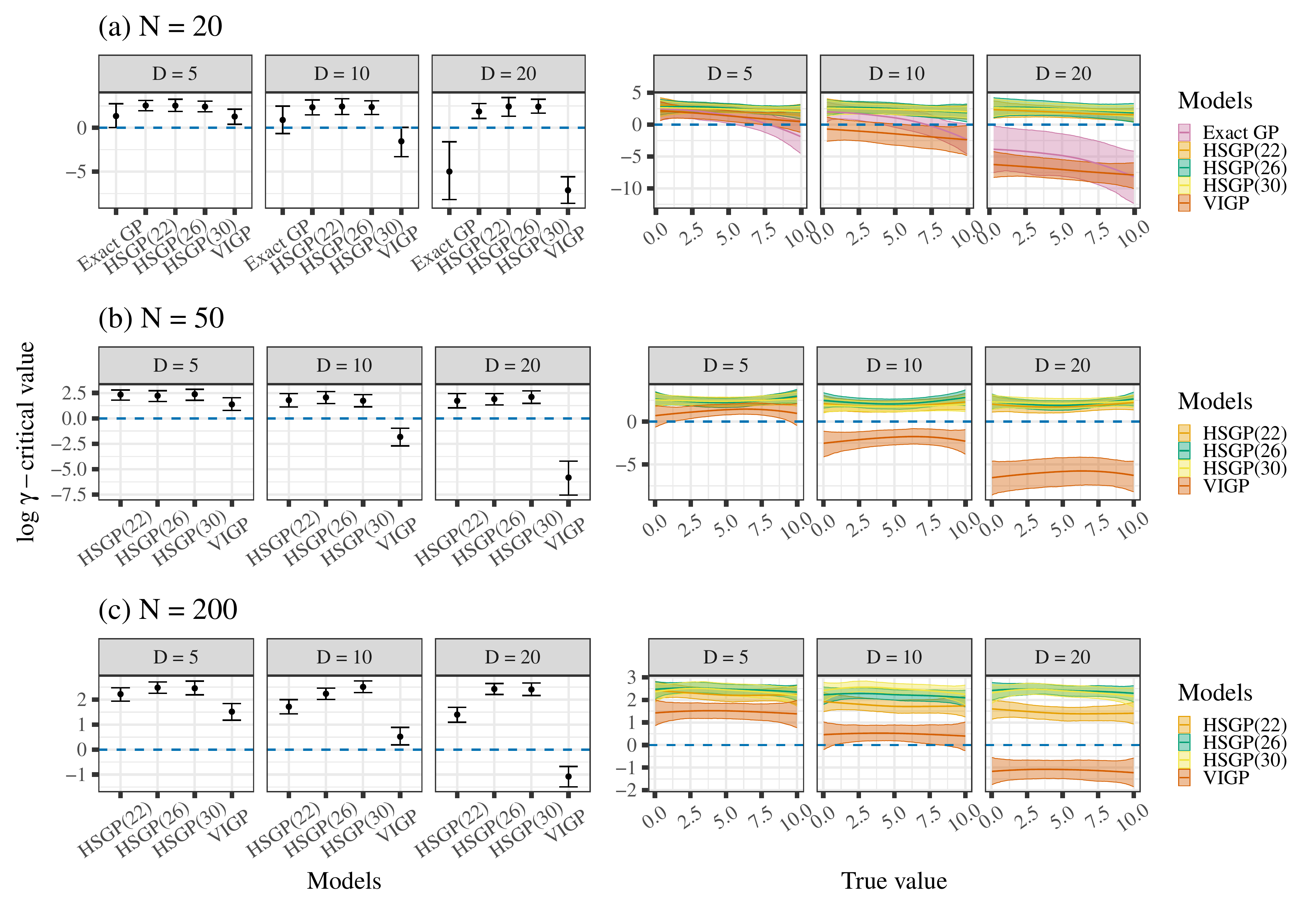}
    \caption{\textit{Squared exponential scenario: log $\gamma$ scores offset by the 95\% confidence threshold for all the fitted models. The behavior of scores across true latent $x$ values are shown in the right-hand panel. The blue dashed line denotes the threshold to reject uniformity, that is, models with values less than $0$ are miscalibrated. The HSGP($M$) shows the HSGPs with their corresponding number of basis functions.}}
    \label{fig:se-log-gamma}
\end{figure}

In Figure \ref{fig:se-ecdf-n20-d20}, we show the rank ECDFs obtained from the exact GPs, HSGPs, and VIGPs for the SE simulation scenario ($N= 20$ and $D=20$). The results indicate that the latent variable estimates obtained from the HSGPs are consistently well calibrated. 
In contrast, for exact GPs, we see miscalibrations for some latent variables shown by their corresponding rank sample ECDFs lying outside the threshold region. In case of VIGPs, the instances of miscalibrations are even more severe. The rank ECDFs for Matern 3/2 and 5/2 scenarios are reported in supplementary materials Section E (Figures S12 and S13) where the results are same as the SE scenario. 

To streamline calibration checking across simulation scenarios , we analyze the log $\gamma$ scores using a multi-level model \citep{burkner_brms_2017} described in supplementary materials Section A. In Figure \ref{fig:se-log-gamma}, we show the predicted log $\gamma$ scores of the different GP models under the SE simulation scenario. This is equivalent to summarizing the different cases of rank ECDFs across all simulation scenarios. 
\begin{figure}[!ht]
    \centering
    \includegraphics[width = \linewidth]{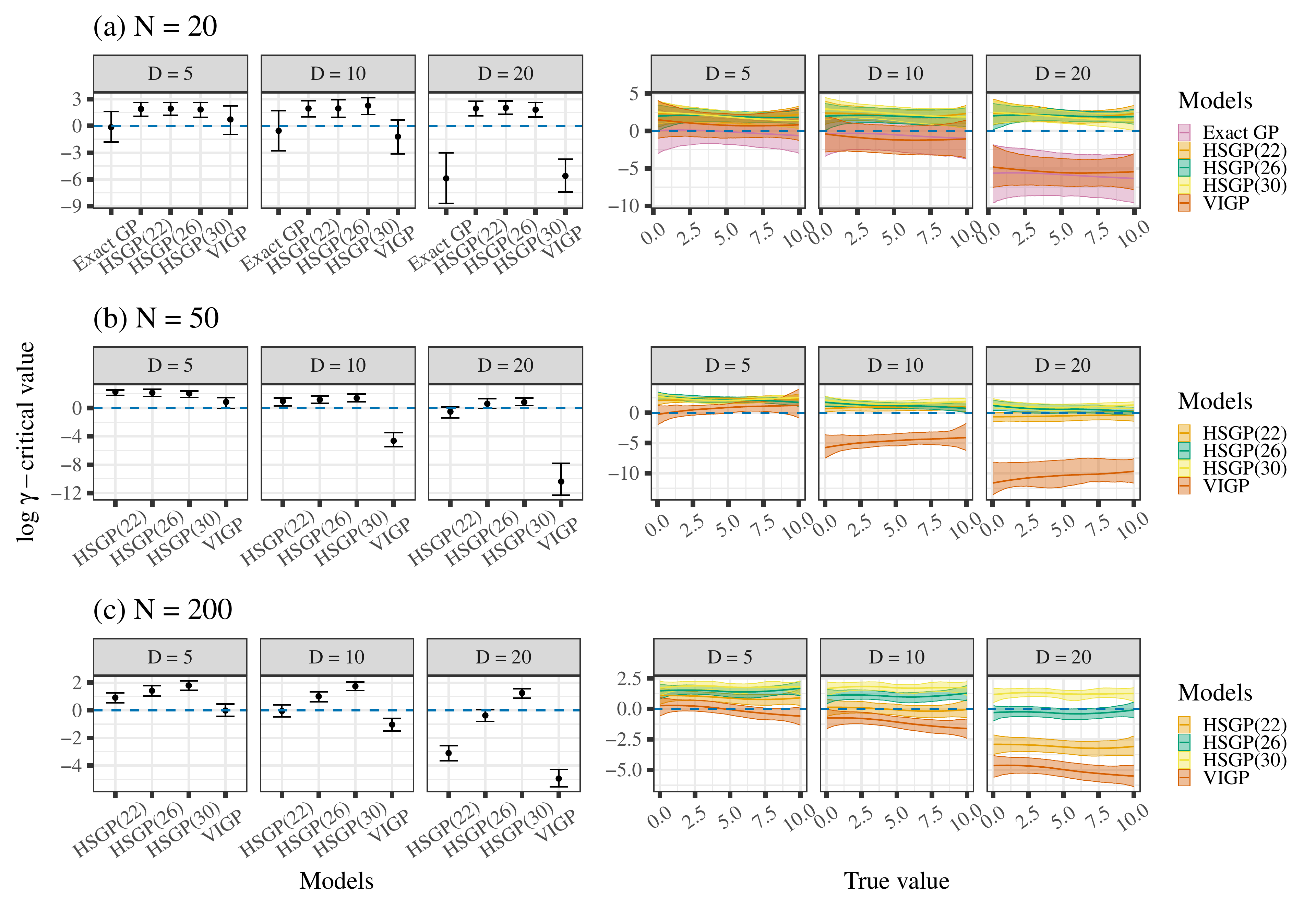}
    \caption{\textit{Squared exponential scenario (highly varying $\rho$): log $\gamma$ scores offset by the 95\% confidence threshold for all the fitted models. The behavior of scores across true latent $x$ values are shown in the right-hand panel. The blue dashed line denotes the threshold to reject uniformity , that is, models with values less than $0$ are miscalibrated. The HSGP($M$) shows the HSGPs with their corresponding number of basis functions.}}
    \label{fig:se-widerho-log-gamma}
\end{figure}
For all the choices of $N$ and $D$, the latent variable estimates from HSGPs are consistently well calibrated. By comparison, the exact GPs fail the calibration test for $D = 20$ as their log $\gamma$ scores remain consistently less than the critical value corresponding to the 95\% confidence threshold. This can be primarily attributed to the computational issues of exact GPs for latent variable inputs as seen through MCMC convergence diagnostics in Section \ref{sec-model-conv} along with supplementary materials Section C and D. The VIGPs shows miscalibrations especially for higher output dimension cases throughout all the choices of $N$. Additionally, exact GPs exhibit worse calibrations (decreasing log $\gamma$ scores) when estimating larger values of the latent inputs. This issue directly causes problems in estimating latent values closer to the boundary. VIGPs show similar behavior for lower sample sizes $N = 20$ although not as strong as exact GPs. The HSGPs are able to overcome and rectify this behavior. For the special scenario of SE data with highly variable length-scales as shown in Figure \ref{fig:se-widerho-log-gamma}, VIGPs and HSGPs with small number of basis function struggle to pass the calibration tests, especially for $N=200$. The miscalibrations are however alleviated with a higher number of basis functions in HSGPs. The log $\gamma$ results for the Matern 3/2 and 5/2 simulation scenarios are presented in supplementary materials Section E (Figures S14 and S15).

\subsection{Latent variable estimation}\label{sec-latent-input-est}
For all of the simulation scenarios discussed in Section \ref{sec-sim-data}, we evaluate the estimation capabilities of the different fitted models specified in Section \ref{sec-model-spec}. We show model performance in terms of absolute bias 
$|\text{Bias}(x_{post}, x_{true})|$ of the latent posterior estimates $x_{post}$ with respect to their true values $x_{true}$ as well as $\text{SD}(x_{post})$ as a measure of posterior sharpness. Since mean-field VI is known to underestimate the posterior SD \citep{advi2017}, we avoid using a metric like RMSE that favors (too) narrow posteriors, thus providing misleading interpretations if used as a sole metric.
\begin{figure}[!ht]
    \centering
    \includegraphics[width = \linewidth]{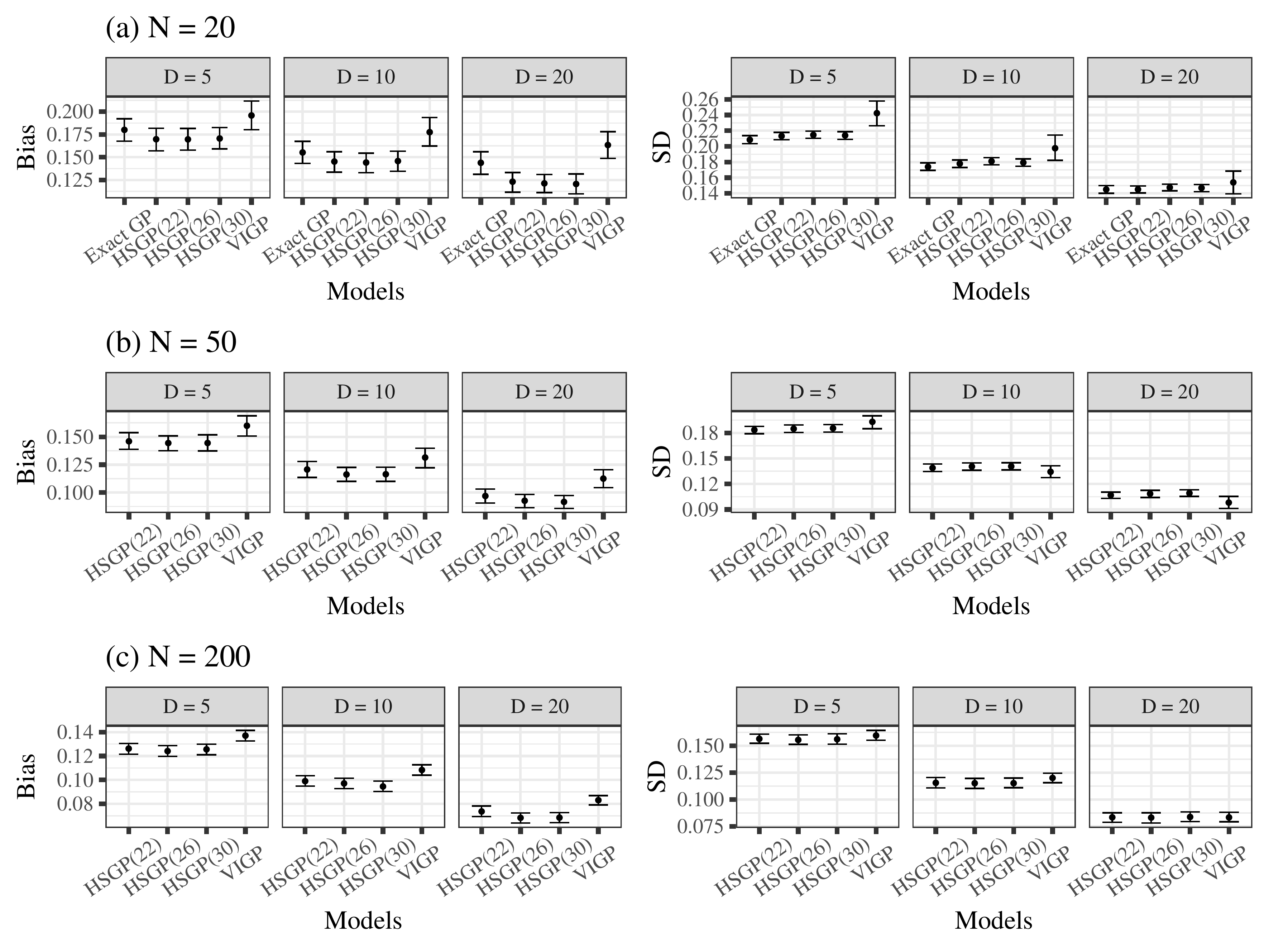}
    \caption{\textit{Squared exponential scenario: posterior bias and SD on recovery of latent inputs for all fitted models. The HSGP($M$) shows the HSGPs with their corresponding number of basis functions.}}
    \label{fig:se-latentx}
\end{figure}
We summarize our findings using the multi-level model \citep{burkner_brms_2017} described in supplementary materials Section A for both posterior bias and SD. The SE simulation scenario presented in Figure \ref{fig:se-latentx} shows that, in most cases, HSGPs have lower bias compared to exact GPs and VIGPs. The VIGPs consistently show the highest bias. Overall, for the SE data scenario, we see on average about 11\% (for higher sample sizes) to about 18\% (for lower sample sizes) decrease in posterior bias between VIGPs and HSGPs (with $M=22$ basis functions). Results for the posterior SD are more mixed, with VIGPs showing higher SDs for small sample sizes and about the same SDs for larger sample sizes. Specifically, HSGPs overcome the computational challenges (in MCMC convergence and posterior uncertainty calibration) of exact GPs to provide better latent variable estimation accuracy.
\begin{figure}[!ht]
    \centering
    \includegraphics[width = \linewidth]{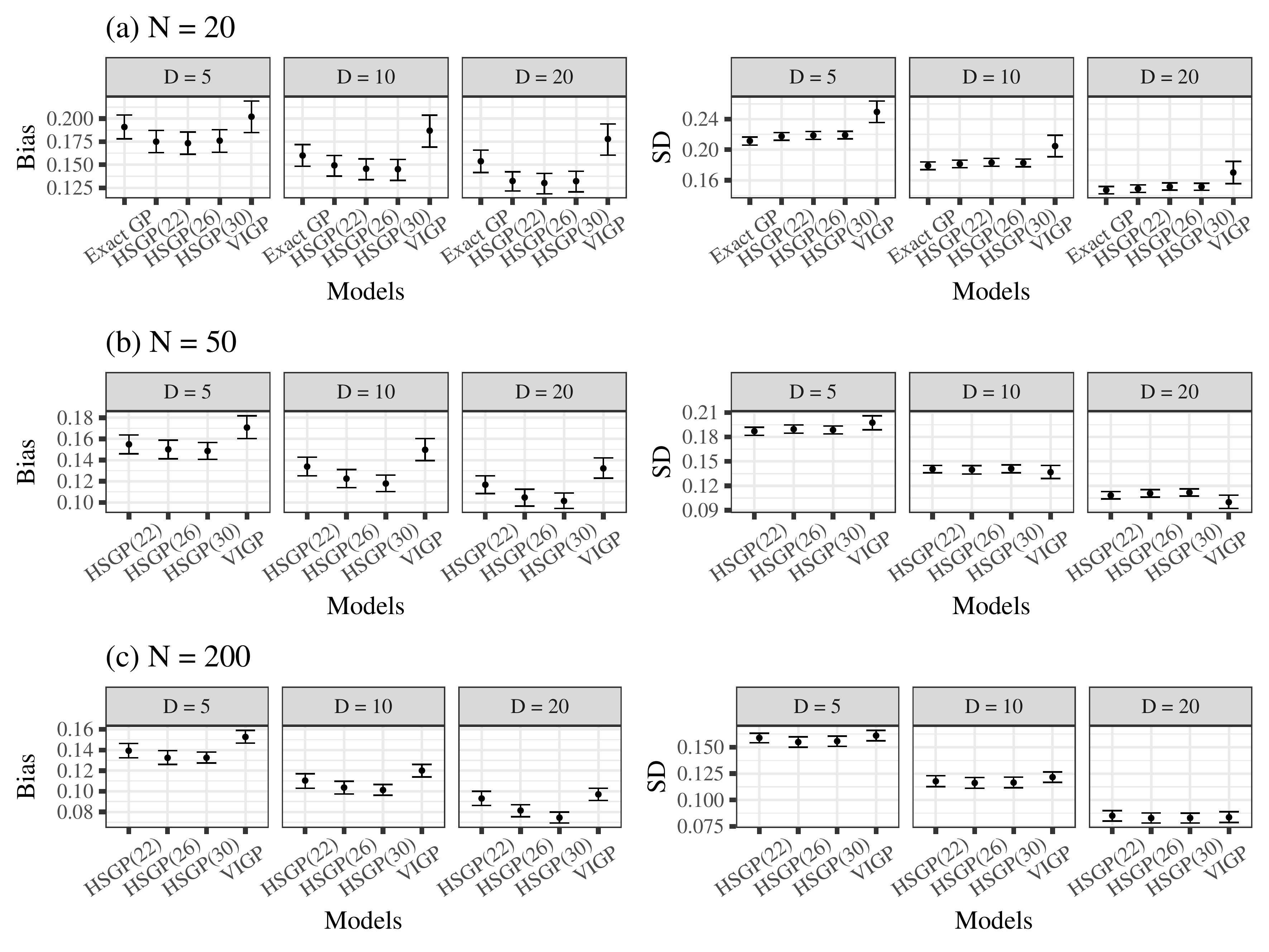}
    \caption{\textit{Squared exponential scenario (highly varying $\rho$): posterior bias and SD on recovery of latent inputs for all fitted models. The HSGP($M$) shows the HSGPs with their corresponding number of basis functions.}}
    \label{fig:se-widerho-latentx}
\end{figure}

In Figure \ref{fig:se-widerho-latentx} we show the more challenging simulation scenario. Here, the data generating process had a higher amount of variability in length-scales across the output dimensions. We observe a stronger difference in posterior bias between VIGPs and HSGPs (in favor of HSGPs). We hypothesize that HSGPs are more strongly favored here as they allow for varying hyperparameters across dimensions, while the VIGPs (at least their Pyro implementation) only support single hyperparameters across all output dimensions. For example, a single length scale parameter $\rho$ is optimized for all $D$ output dimensions. While this is a limitation of the current VIGP implementation, we compare a favorable case where the GP hyperparameters are fixed to the data generating conditions in the special simulation scenario with $N = 1000$ in Section \ref{sec-simspl-case}. 
 
Among the HSGPs, higher number of basis functions ($M = 26 \text{ } \text{and} \text{ } 30$ compared to $M = 22$) result in lower posterior bias for both the simulation scenarios shown in Figures \ref{fig:se-latentx} and \ref{fig:se-widerho-latentx}. Combining this result with their corresponding model calibration tests, we recommend using $M = 30$ to achieve the highest accuracy for the latent variable estimates. For faster but reliable results, HSGPs with $M=22$ can be used in simpler cases like the SE data scenario. Furthermore, in the remaining scenarios (see Figures S16-S19 in supplementary materials Section F), HSGPs perform consistently better than their competitors. In all the scenarios, each of the fitted models (exact GPs only for $N=20$) show a consistent decrease in posterior bias and SD as the number of output dimensions $D$ and sample size $N$ increases. We report the estimation results for GP hyperparameters in the supplementary materials Section B (followed by additional figures in Section H).

\subsection{Checking robustness of HSGPs with wide priors} \label{sec-simspl-case}
In the above discussed GP simulation scenarios, the hyperparameter priors for exact GPs and HSGPs are aligned with the data generating conditions as a requirement for SBC checks (see Section \ref{sec-model-calib}). Therefore, in those simulation studies, the exact GPs and HSGPs receive a favorable treatment as compared to VIGPs. Hence, to check the robustness of latent variable HSGPs, we perform additional simulations where we apply hyperparameter priors to the HSGPs that are less informed (i.e., wider priors) along with a larger sample size $N = 1000$. We leave exact GPs out of this simulation study due to their computational limitations. 

Keeping the data generating conditions same as discussed in Section \ref{sec-sim-data}, we specify priors for length-scale $\rho \sim \text{Normal}^+(1.5, 0.5^2)$, GP marginal SD $\alpha \sim \text{Normal}^+(3.5, 1^2)$ and error SD $\sigma \sim \text{Normal}^+(1.5, 1^2)$. That is, we shift the prior mean away from the data generating conditions along with a much wider prior variance. A comparison of the data generating conditions and the GP hyperparameter priors are presented in Table \ref{tab:data-prior-compare}. Simulations are carried out for SE, Matern 3/2 and 5/2 covariance functions. Based on the specified length-scale prior and chosen covariance function, the minimum required number of basis functions for HSGPs change according to the empirical relation shown in Eq.\eqref{eqn-m-se}-\eqref{eqn-m-m52}.

\begin{table}[!ht]
     \centering
     \caption{Comparison of GP hyperparameter priors and data generating conditions}
     \begin{tabular}{c c c}
     \toprule
      Hyperparameter & Dgp prior & Model prior \\
  \midrule
 $\rho$ & $\text{Normal}^+(1, 0.05^2)$ & $\text{Normal}^+(1.5, 0.5^2)$ \\
 $\alpha$ & $\text{Normal}^+(3, 0.25^2)$ & $\text{Normal}^+(3.5, 1^2)$ \\
 $\sigma$ & $\text{Normal}^+(1, 0.25^2)$ & $\text{Normal}^+(1.5, 1^2)$ \\
  \bottomrule
     \end{tabular}
     \begin{tablenotes}
        \item \small \textit{Note: The Dgp priors show the true sampling distribution for simulated data and Model priors show the prior distribution for the HSGP parameters.}
     \end{tablenotes}
     \label{tab:data-prior-compare}
 \end{table}

We consider two VIGP variants: VIGP (default) and VIGP (prior). The former is the same VIGP model considered in other simulation scenarios with the default priors used by Pyro. In contrast, for VIGP (prior), we fix the GP hyperparameters to the mean of the true sampling distributions, thereby acting as a highly informed reference. 

\begin{figure}[!ht]
    \centering
    \includegraphics[width = \linewidth]{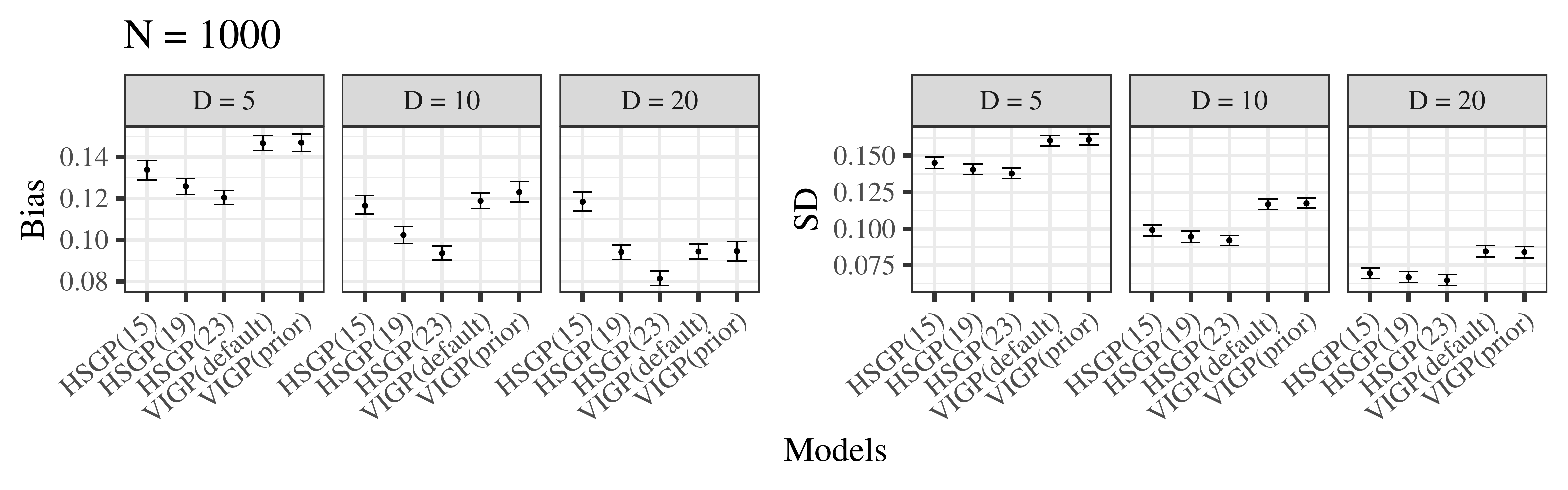}
    \caption{\textit{Squared exponential scenario (special case): posterior bias and SD on recovery of latent inputs for all fitted models. The HSGP($M$) shows the HSGPs with their corresponding number of basis functions. VIGP (default) shows the default Pyro implementation and VIGP (prior) has fixed GP hyperparameters informed by the true sampling distributions.}}
    \label{fig:se-n1000-latentx}
\end{figure}

We summarize the posterior bias and SD using the same multilevel model (described in supplementary materials Section A) used for other simulation scenarios. For the SE scenario in Figure \ref{fig:se-n1000-latentx}, the HSGPs achieve lower bias compared to VIGPs with a higher number of basis functions $M$. This is especially true for higher output dimensions $D = 20$ where we need much higher $M$ than the minimum $M$ suggested by the empirical relation in Eq.\eqref{eqn-m-se}. On the other hand, posterior SDs for HSGPs are consistently lower than VIGPs. Among the VIGP variants, both VIGP (default) and VIGP (prior) performs similarly in terms of latent variable estimation accuracy. That is, fixing the GP hyperparameters for VIGPs to informed values does not result in any improvement in model performance. The results for Matern 3/2 and 5/2 covariance functions (see Figures S20 and S21 in supplementary materials section G) are highly similar, with Matern 3/2 showing even more pronounced differences between HSGPs and VIGPs.

The average runtime per dataset of sample size $N = 1000$ on the computing cluster (details mentioned previously) was about 6.6 hours on average for SE, 11.5 hours for Matern 3/2 and 10.2 hours for Matern 5/2 models. For these specific simulation conditions, the runtime for HSGPs are reasonable compared to exact GPs that are computationally infeasible. Although the HSGP computation times are affected due to wider prior specifications for GP hyperparameters, the model performance in terms of latent variable estimation accuracy remains superior to VIGPs as in other simulation scenarios. In terms of speed, the VIGPs were similarly fast as in other simulation scenarios.

\section{Real-World Case Study} \label{sec-case-study}
We illustrate the application of the HSGPs using a real-world scRNA sequencing data. To that end, we re-analyze cell-cycle data from \cite{mahdessian_spatiotemporal_2021}. Our aim is to estimate the underlying cell ordering known as "pseudotime". The dataset comprises of spliced single-cell messenger RNA (or mRNA) expression profiles of cells along the cell cycle. Each cell is additionally labeled with its own approximated time in the cell cycle process, i.e. the experimental time called "cell hours". We choose this dataset because it covers a cyclic process going through three distinct biological phases (G1, G2M and S-ph) depicting substantial variation in gene expressions. Additionally, the provided cell hours is a continuous temporal sequence that can be utilized as a prior guide to estimate pseudotime. For the purpose of this real-world case study, we use the data from all the cells along with a selection of 12 influential genes (see supplementary materials Section I for the list) based on the recommendations of \cite{mahdessian_spatiotemporal_2021}. In other words, each sample point corresponds to a single cell and each output dimension corresponds to a single gene, with the value being the gene expressions per cell. Thus, the analyzed data comprises of $N = 960$ observations and $D=12$ output dimensions. 
\begin{figure}[!ht]
    \centering
    \includegraphics[width = \linewidth]{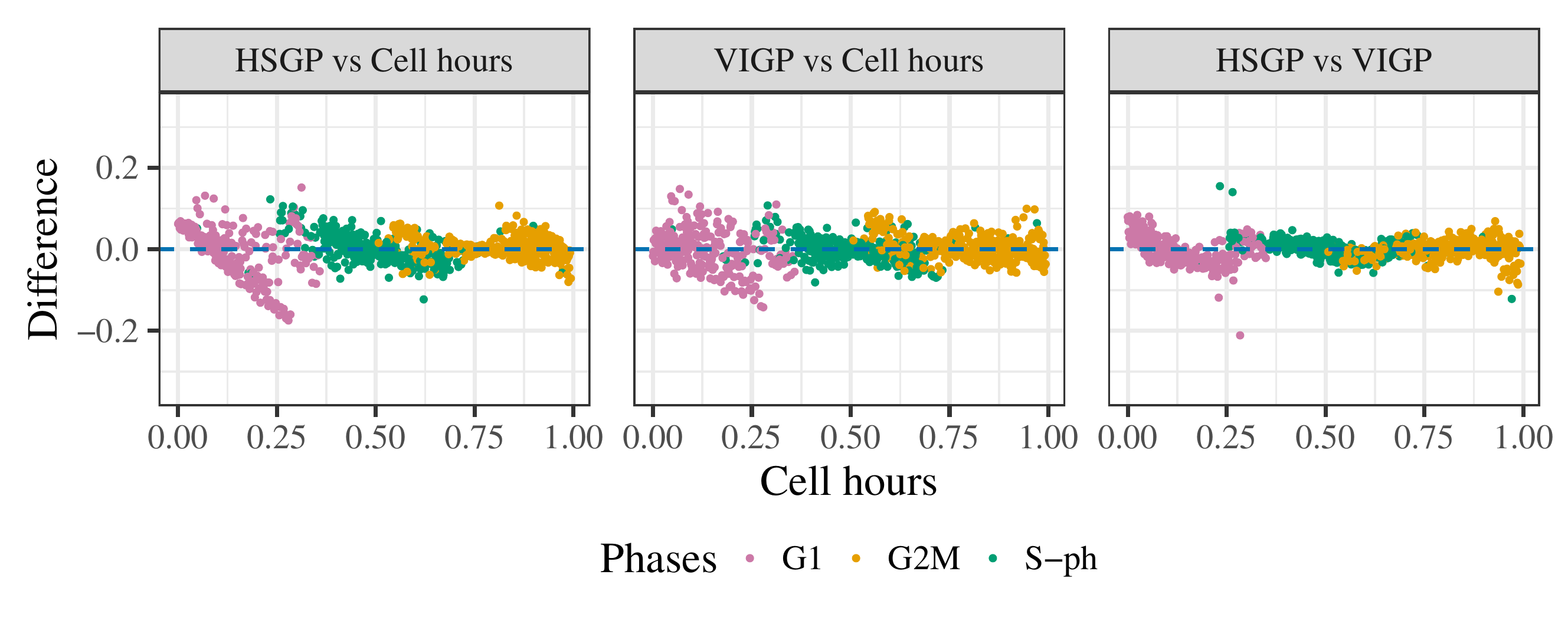}
    \caption{\textit{(a) Difference of HSGP pseudotime estimates from the prior cell hours; (b) Difference of VIGP pseudotime estimates from the prior cell hours; (c) Difference of HSGP and VIGP pseudotime estimates. Both models assume an SE covariance function. The different colors denote the three distinct biological phases (G1, G2M and S-ph).
    }}
    
    \label{fig:case-study-diff}
\end{figure}
We use cell hours in the context of this specific data as $\tilde{x}$ for our latent pseudotime $x$. Both $\tilde{x}$ and $x$ are real numbers with values ranging between 0 and 1. For our prior measurement SD which is unknown, we choose $s = 0.03$, keeping it proportional to our choices in the simulation studies in Section \ref{sec-sim-data} and \ref{sec-model-spec}.

We fit a HSGP and VIGP with SE covariance function. We specify the priors for length-scale $\rho \sim \text{Normal}^+(0.4, 0.1^2)$, GP marginal SD $\alpha \sim \text{Normal}^+(14, 3.5^2)$, and error SD $\sigma\sim \text{Normal}^+(7, 3.5^2)$. The prior specifications are based on the similar real-world case study in \cite{mukherjee2025dgplvmderivativegaussianprocess}. Based on the $\rho$ prior and the $(0, 1)$ range of the input space, we use the minimum number of basis function $M = 6$ (see Eq.\eqref{eqn-m-se}). For the VIGP, we again resort to default priors since custom priors are not supported. Additionally, based on the $N = 1000$ simulation study results in Section \ref{sec-simspl-case}, the default implementation performs equally well compared to fixed hyperparameter settings. For the number of inducing points, we follow \cite{ahmed_grandprix_2019} where $10$ inducing points are shown to be sufficient for pseudotime estimation. Moreover, keeping the number of inducing points to $10$ similar to our simulation studies allows us to avoid compromising the numerical stability of VIGPs.

As in any real-world latent variable problem, we lack knowledge of the ground truth for the latent variable. Thus, we show the deviation of the posterior pseudotime $x$ estimates from the prior cell hours $\tilde{x}$. The results from our simulation studies provide evidence that this shift is indeed in the direction of the ground truth. The pseudotime deviations from cell hours for both HSGP and VIGP are presented in Figure \ref{fig:case-study-diff}. The first two sub-figures show the estimated pseudotime deviations from the HSGP and VIGP, respectively, plotted against the prior cell hours. The third sub-figure show the difference between the estimated pseudotimes from these two models. For some cells, prior-posterior differences are on average about 10\% of the total cell hours time scale for both HSGP and VIGP. Comparing their two pseudotime estimates, we see differences of about 5\% (with some outlying cells).
\begin{figure}[!ht]
    \centering
    \includegraphics[width = \linewidth]{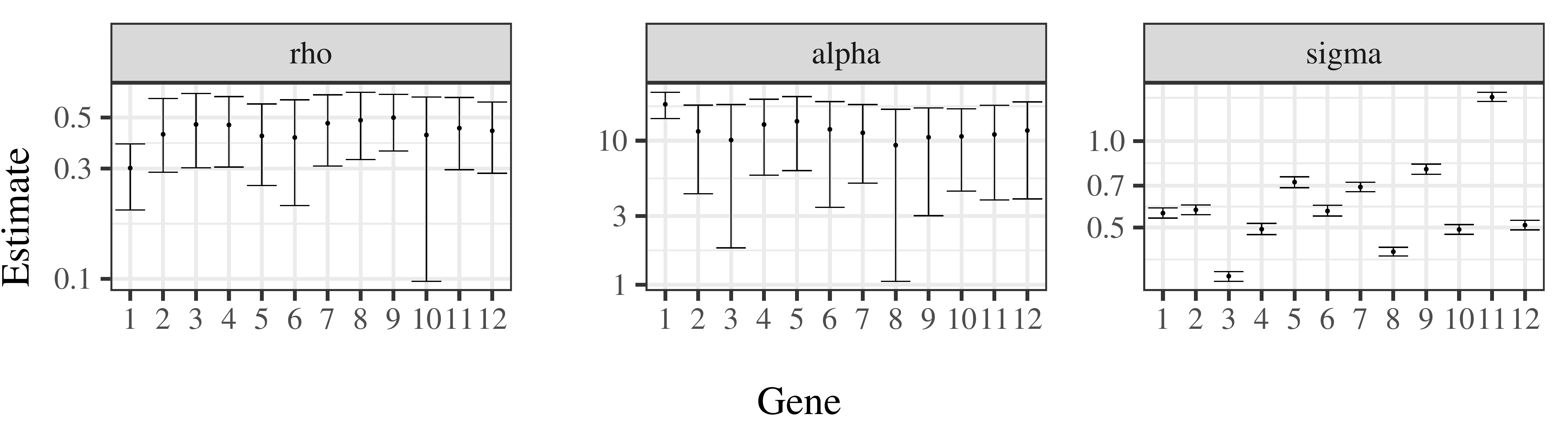}
    \caption{\textit{Hyperparameter estimates for the HSGP with SE covariance function for cell cycle data. Points indicate posterior means and the point ranges indicate 95\% CIs for each hyperparameter per output dimension.}}
    \label{fig:case-study-pars}
\end{figure}

In Figure \ref{fig:case-study-pars}, we show the posterior estimates of the HSGP hyperparameters along with their posterior 95\% CIs. A key observation here is the strong differences in the hyperparameter estimates across the genes (output dimensions). This confirms the importance of assuming gene-specific hyperparameters while modeling these biological processes. Modeling varying GP hyperparameters, while supported in our HSGPs, is not possible yet for the (pyro implemented) VIGPs, so we do not include their estimates in Figure \ref{fig:case-study-pars}.

The runtime for the HSGP was approximately 16 minutes on an Apple M4 chip (2 cores in parallel) with 24 GB working memory. Same model fitted using the cluster computing resources detailed in Section \ref{sec-model-spec} took about 58 minutes. In comparison, fitting the VIGP took about 3 minutes using the same cluster computing resources.

\section{Discussion} \label{sec-discuss}
We generalize Hilbert space approximations for Gaussian processes \citep[HSGPs; ][]{solin_hilbert_2020, riutort-mayol_approx_gps_2022} to develop a scalable class of latent variable models for high-dimensional multi-output data. Our HSGPs account for both varying hyperparameters across output dimensions and correlations between output dimensions. They reduce the steep computational complexity of $O(N^3D + ND^2)$ for exact multi-output GPs to $O(NMD+ND^2)$ for a dataset of $N$ samples and $D$ output dimensions, where $M$ is the number of basis functions of the HSGP. Since $M$ is typically much smaller than $N$, using HSGPs reduce computation times drastically. Through extensive simulations, we investigated the statistical properties of latent GPs in terms of model convergence, uncertainty calibration, and estimation accuracy. In all of the above criteria, the HSGPs performed better than both exact GPs and approximate GPs based on inducing points and variational inference (VIGPs)\citep{titsias_variational_2009, ahmed_grandprix_2019}. While VIGPs are faster yet, they provide limited modeling capabilities and are outperformed by HSGPs in terms of model calibration and estimation accuracy. Moreover, even with less informed hyperparameter priors (wide priors), the HSGPs achieve higher accuracy than VIGPs. Although HSGPs are comparably slower, they are reasonably fast even for relatively large datasets such as the one investigated in our real-world case study. This stands in sharp contrast to exact GPs, which would have taken at least several hours, if not more, to fit the same data.

\subsection{Limitations and future research}
For GPs with Matern class of covariance functions, the length-scale $\rho$ is directly related to the scale of the input space. In latent variable GPs, this implies strong dependencies between the prior on $\rho$, the range of latent $x$, and the prior measurement SD $s$ (see Sections \ref{sec-methods} and \ref{sec-model-spec}). As a result, specifying all three independently may lead to convergence issues and overall bad model behavior. Future research should investigate these dependencies more closely to provide better guidelines for prior specification in latent variable GPs.

With regard to the correlation structure $C$ across output dimensions (see Section \ref{sec-multiout}), we make the choice of modeling it on the level of the GP functions $f$ rather than on the level of observations $y$. It is not always apparent which of these two options
is preferable for a given application. For example, in spatial statistics, correlations between discrete locations are also either modeled on the mean (similar to the GP functions in our case) level, e.g., via conditional autoregressive (CAR) structures, or on the observation level, e.g., via spatial autoregressive (SAR) structures \citep{wall2004close}. In this paper, we apply our methods to RNA gene expressions for cell cycle data, where the genes are known to be correlated at the biological process level \citep{Riba2022cellcycle}. Thus, specifying across-dimension correlations on the GP functions appear sensible here. For other applications, it may be preferable to specify the correlations at the observational level, or even jointly at both levels. 

Another consideration concerning $C$ is whether to model it as input dependent or input independent. In this paper, we chose it to be input independent, thus estimating the full correlation matrix directly. As an alternative, an input dependent formulation could be achieved through another GP that models the correlations across output dimensions within each observation (similar to a cross-covariance structure shown in \cite{Gneiting2010crosscovariance}). In this regard, it would further be interesting to study Hilbert space approximations also for this across-output dimension GP, especially when the number of dimensions is large enough for exact GPs to become computationally impractical. We leave these extensions of our framework to future research. 

Lastly, an important generalization of latent variable GPs concern the inclusion of derivative information, which can further increase estimation accuracy \citep{mukherjee2025dgplvmderivativegaussianprocess}. However, such methods have so far only been developed for exact GPs, which hamper their real-world applicability. Extending HSGPs to include derivative information will likely provide a scalable solution. However, such an extension requires spectral densities for composite covariance functions, which to our knowledge is yet to be derived.

\section*{Code availability}
The code for the model development, simulation studies as well as the results can be found here: \url{https://github.com/Soham6298/Latent-variable-HSGPs}

\section*{Acknowledgments}
We acknowledge the valuable insights provided by all group members. P.B further acknowledges support of the Deutsche Forschungsgemeinschaft (DFG, German Research Foundation) via the Collaborative Research Center 391 (Spatio-Temporal Statistics for the Transition of Energy and Transport) – 520388526. The authors thank the International Max Planck Research School for Intelligent Systems (IMPRS-IS) for supporting S.M.

\bibliographystyle{apalike}
\bibliography{refs}

\newpage
\section*{Supplementary materials} \label{App}

\section*{A: Methods related to simulation study} \label{Sup:sim-summary-methods}
\subsection*{Summarizing results} \label{Supp:summary-methods}
To evaluate the accuracy of estimating latent variables, we compare posterior samples of the latent input variable $x$ denoted by $x_{post}$ from each model with their respective ground truth values denoted by $x_{true}$. Using $\text{Bias}(x_{post}, x_{true})^2$ and $\text{SD}(x_{post})$, we look at the bias-SD trade-off in estimating the latent variables. We compute the posterior bias and SD from all fitted models for each sample size ($N= 20, \text{ }50\text{ } \text{and}\text{ } 200$) and output dimension ($D= 5, \text{ }10\text{ } \text{and}\text{ } 20$). We prefer models that provide both low bias indicating posterior mean estimates close to the ground truth as well as lower posterior SD indicating high precision. The reliability of posterior SD estimates depend on our models being well calibrated.

To analyze the results from our experiments, we use a multilevel analysis of variance model (ANOVA) fitted with brms \citep{burkner_brms_2017}, which disentangles the various components of our simulation study design. With $\mu_{resp}$ and $\sigma_{resp}$ being the mean and SD of our response variable, we use
\begin{equation}
    \begin{aligned}
        \mu_{resp} &= X\beta + Zb+\sum_{\tau}s_{\mu_\tau}(t_\tau) &\\
        \sigma_{resp} &= X\eta + Zu+\sum_{\tau}s_{\sigma_\tau}(t_\tau)
    \end{aligned}
\end{equation}
where $\beta, \text{ }b$ (for $\mu_{resp}$) and $\eta, \text{ } u$ (for $\sigma_{resp}$) are coefficients at the population and group levels with $X$ and $Z$ being the corresponding design matrices. The $s_\tau(t_\tau)$ terms denote smooth functions over covariates $t$ fitted via splines. We use these models to summarize the results from our experiments. In the simulation studies, the population level design matrix $X$ contains covariates representing the approximation method and number of output dimensions along with their interaction terms. Based on our experimental setup, the approximation methods include the exact model, HSGPs (with three choices of basis functions) and VIGPs, thus creating a five level factor variable. We use a three level factor variable depicting the $5$, $10$, and $20$ output dimensions. Through $Z$ we account for the group-level dependency structure in the response induced by fitting multiple models to the same simulated dataset. We include a random intercept over datasets as well as corresponding random slopes of the approximation method choices. Lastly, to capture the non-linear relation of the response to the ground truth $t$, we introduce thin-plate regression spline terms $s_{\mu_\tau}(t_\tau)$ and $s_{\sigma_\tau}(t_\tau)$. The spline terms accounts for any non-linear relationship of the response with respect to the true parameter values $t$. We use the log $\gamma$ scores as our response to summarize the model calibration results. To summarize the results of latent variable estimation, we subsequently specify posterior bias and SD as the response.

\subsection*{SBC} \label{Supp:sbc}
Using simulation-based calibration (SBC), we test model calibration for estimating latent inputs $x$. We start with a model, say, $\mathcal{M}_0$. We then generate $J$ datasets $y^{(j)}, j = 1,\dots,J$ each of size $N$ from the data generating process that exactly aligns with the model $\mathcal{M}_0$. In other words, each individual dataset $y^{(j)}$ is generated based on a corresponding model parameter draw $x_0^{(j)}$ from its prior distribution $p(x)$. We sample from the posterior approximation by fitting $\mathcal{M}_0$ to each of the datasets $y^{(j)}$ thus resulting in $J$ fitted models $\mathcal{M}^{(j)}$ with respective posteriors $p(x \mid y^{(j)})$ each having $H$ posterior draws $x^{(j, h)}$. Using $x_0^{(j)}$ as the ground truth, we then calculate a rank statistic for each univariate posterior quantity $\mathbf{h}_p(x)$ for a specific parameter by counting the number of posterior draws $\mathbf{h}_p(x^{(j, h)})$ that are smaller than $\mathbf{h}_p(x_0^{(j)})$. The rank statistic $R^{(j)}$ for the model $T^{(j)}$ is then given as 
\begin{equation} \label{eqn-sbc-rank}
    R^{(j)} = \sum_{h=1}^H \mathbb{I}[\mathbf{h}_p(x^{(j, h)}) < \mathbf{h}_p(x_0^{(j)})]
\end{equation}
The distribution of these single rank-value per model taken together across all $J$ models is a discrete uniform distribution if the approximate posteriors correspond to the true posteriors. Using this property, we assess the correctness of the posterior approximations by testing the rank distribution for uniformity. If the rank distribution departs from uniformity, it indicates a problem in the data generating process, model implementation, the posterior approximations or a combination of these. 

We check SBC by computing the confidence bands for the empirical cumulative distribution function (ECDF) of the rank distribution and visualizing it under the assumption of the uniformity. An alternative, however is based on the probability of $\gamma$ observing the most extreme point on the ECDF under the assumption of uniformity. The test statistic is given by 
\begin{equation} \label{eqn-gamma-stat}
    \gamma = 2 \quad \min_{j \in \{1, \dots, J+1\}}(\min\{\text{Bin}(R_j \mid J, z_j), 1 - \text{Bin}(R_j-1 \mid J, z_j)\}),
\end{equation}
where $z_j$ is the expected proportion of ranks below $j$ such that $z_j = \frac{j}{J+1}$, $R_j$ is the actual empirical ranks below $j$, $\text{Bin}(R_j \mid J, z_j)$ is the CDF of the Binomial distribution with $J$ trials and the probability of success evaluated at $R$. The calculated $\gamma$ scores (presented on the logarithmic scale) are then compared to a threshold value at which to reject uniformity. The log $\gamma$ scores are advantageous in summarizing large number of parameters, different models as well us various simulation conditions. Thus, we prefer evaluating model calibration using the log $\gamma$ scores in our case.

\section*{B: GP hyperparameters estimation}\label{Supp:hyperparams-est}
We show the estimation accuracy for hyperparameters of the exact GPs and HSGPs for all the simulation scenarios.The VIGP doesn't allow varying hyperparameter estimation for different output dimensions. Moreover, model inference using VI only provides a posterior point estimate for each of the hyperparameters and are thus excluded from this part of model evaluation comparison. We present the model evaluation summary using RMSE which combines the model-specific effects on posterior bias and SD. 

\begin{figure}[!ht]
    \centering
    \includegraphics[width = \linewidth]{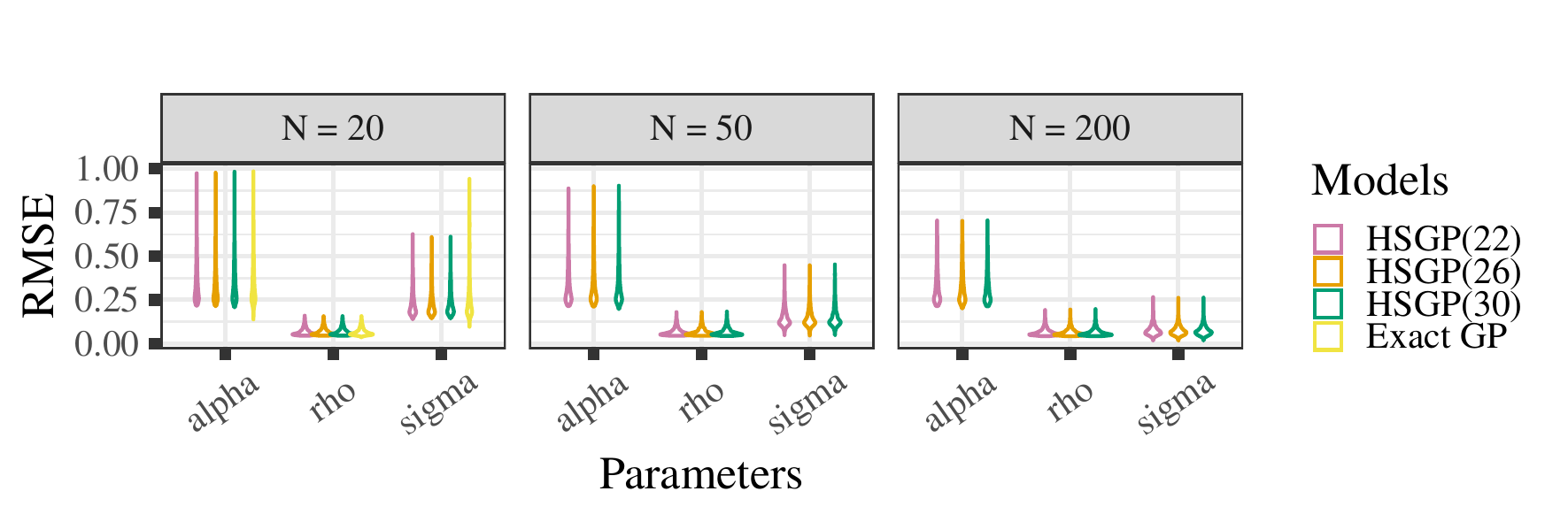}
    \caption{\textit{Squared exponential scenario: RMSE on recovery of GP hyperparameters for exact GP and HSGP fitted models. The HSGP($M$) shows the HSGPs with their corresponding number of basis functions.}}
    \label{fig:se-hyperparams}
\end{figure}
\begin{figure}[!ht]
    \centering
    \includegraphics[width = \linewidth]{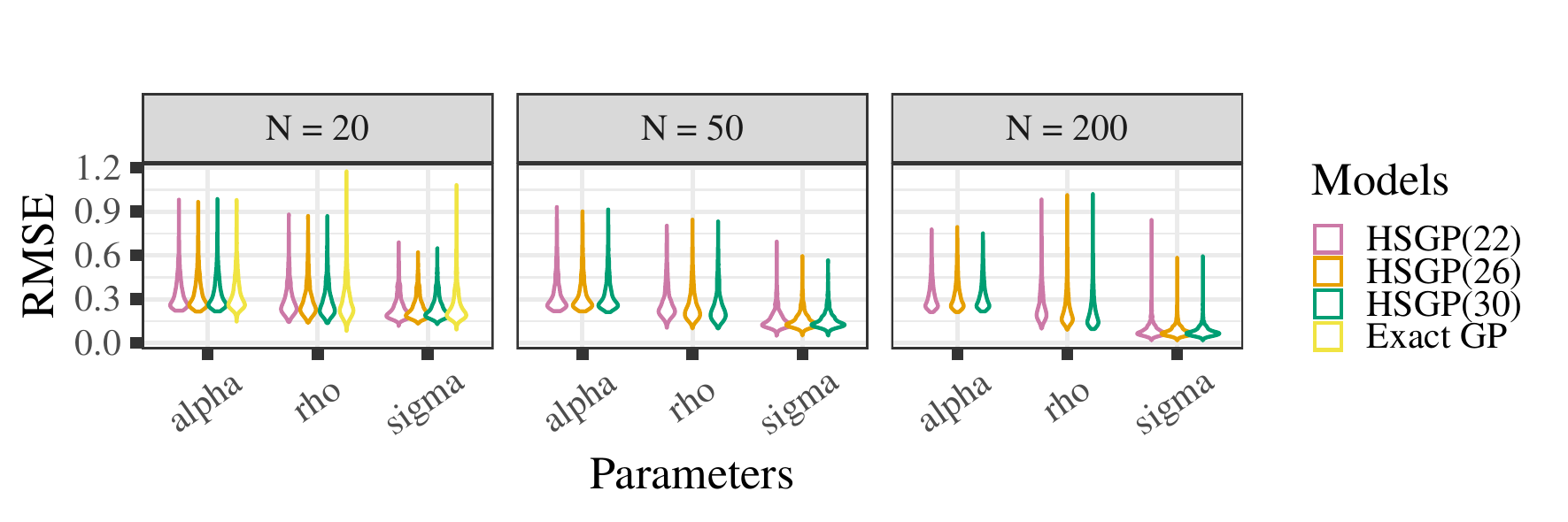}
    \caption{\textit{Squared exponential scenario (highly varying $\rho$): RMSE on recovery of GP hyperparameters for exact GP and HSGP fitted models. The HSGP($M$) shows the HSGPs with their corresponding number of basis functions.}}
    \label{fig:se-widerho-hyperparams}
\end{figure}
\begin{figure}[!ht]
    \centering
    \includegraphics[width = \linewidth]{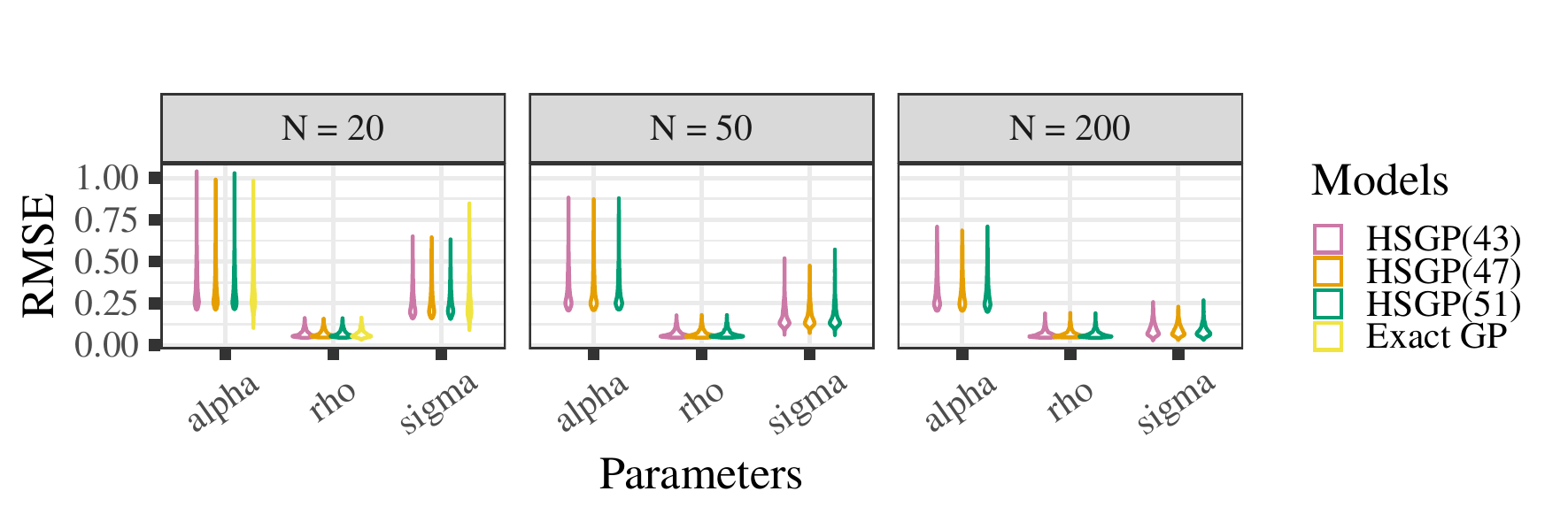}
    \caption{\textit{Matern 3/2 scenario: RMSE on recovery of GP hyperparameters for exact GP and HSGP fitted models. The HSGP($M$) shows the HSGPs with their corresponding number of basis functions.}}
    \label{fig:m32-hyperparams}
\end{figure}
\begin{figure}[!ht]
    \centering
    \includegraphics[width = \linewidth]{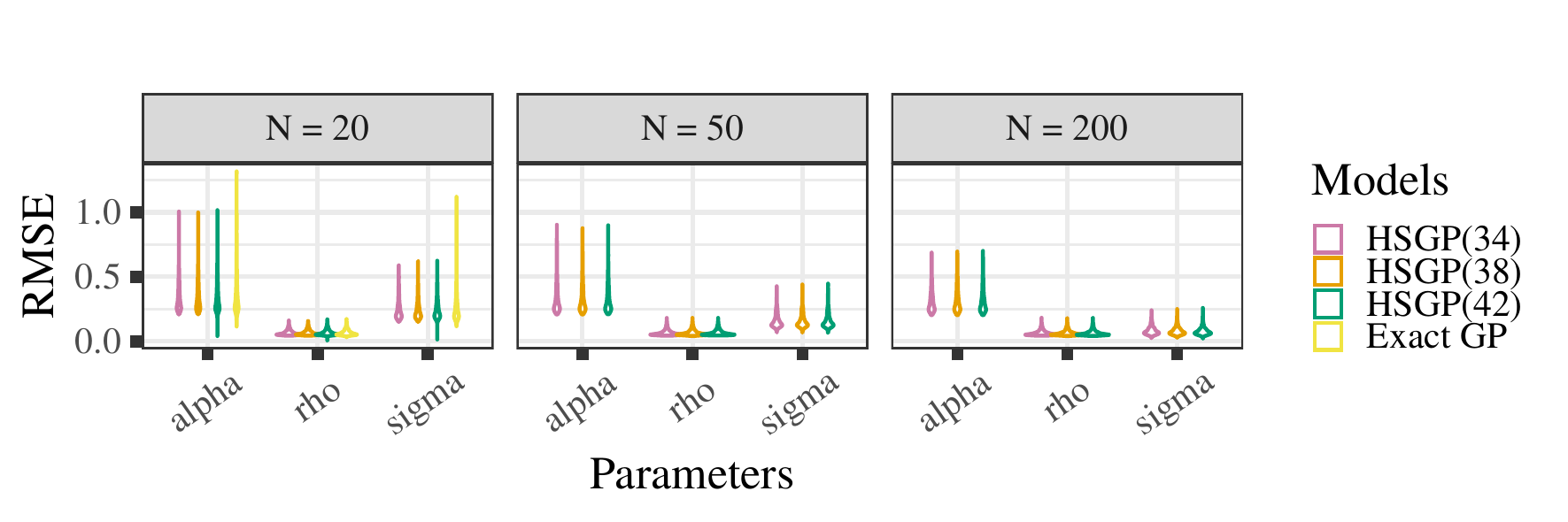}
    \caption{\textit{Matern 5/2 scenario: RMSE on recovery of GP hyperparameters for exact GP and HSGP fitted models. The HSGP($M$) shows the HSGPs with their corresponding number of basis functions.}}
    \label{fig:m52-hyperparams}
\end{figure}
\begin{figure}[!ht]
    \centering
    \includegraphics[width = \linewidth]{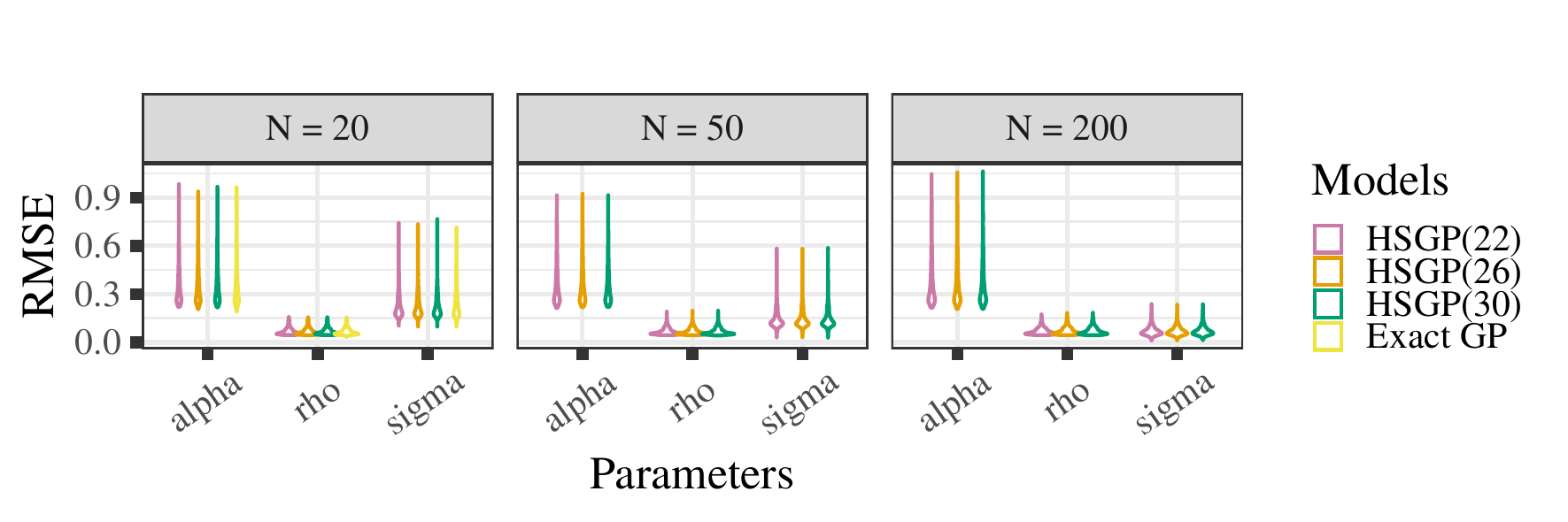}
    \caption{\textit{Periodic data scenario (lower oscillations): RMSE on recovery of GP hyperparameters for exact GP and HSGP fitted models. The HSGP($M$) shows the HSGPs with their corresponding number of basis functions.}}
    \label{fig:per-hyperparams}
\end{figure}
\begin{figure}[!ht]
    \centering
    \includegraphics[width = \linewidth]{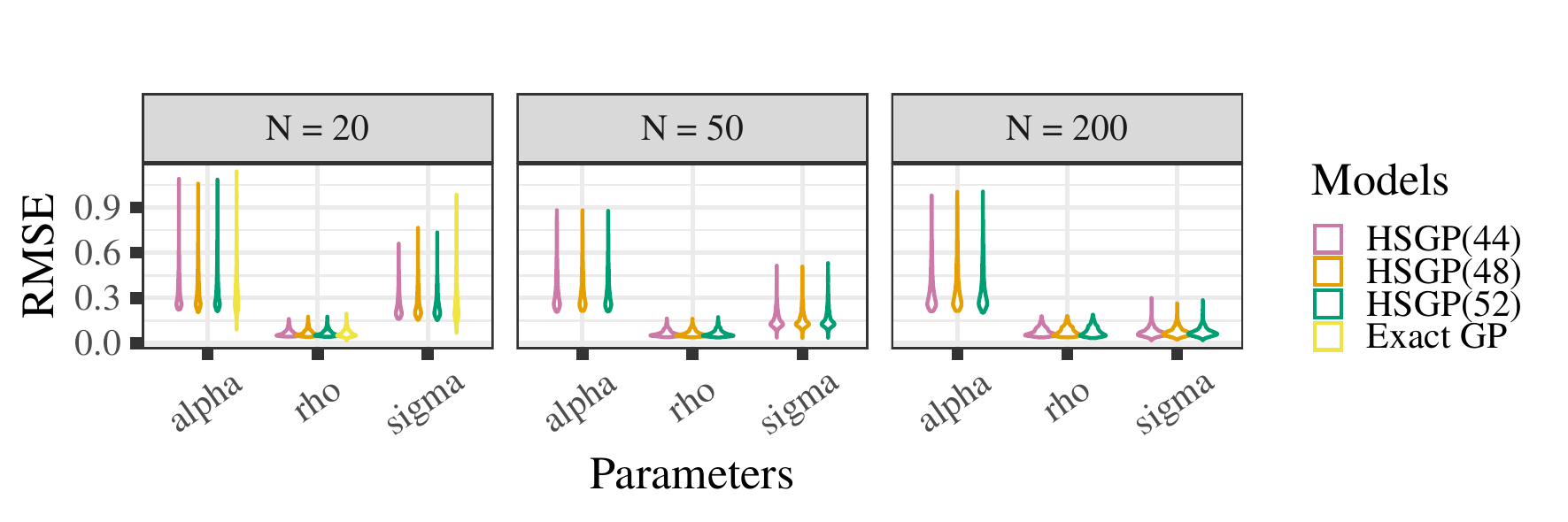}
    \caption{\textit{Periodic data scenario (higher oscillations): RMSE on recovery of GP hyperparameters for exact GP and HSGP fitted models. The HSGP($M$) shows the HSGPs with their corresponding number of basis functions.}}
    \label{fig:per-lowrho-hyperparams}
\end{figure}

We present the RMSE for each model hyperparameter under the SE data scenario in Figure \ref{fig:se-hyperparams} where we see how the decreasing effect on the RMSE of GP marginal SD $\alpha$ and error SD $\sigma$ as the sample size $N$ increases. Since the results were qualitatively same for all the different choices of output dimensions $D$, we decided to showcase only for the $D = 20$. Interestingly, RMSE for the length-scale $\rho$ stays similar across all sample size choices. We can thus be sure about the reliability of HSGPs in recovering true hyperparameter values for all kinds of data scenarios. In the more challenging case of the highly variable length-scale, we see in Figure \ref{fig:se-widerho-hyperparams} that the RMSE for the length-scale $\rho$ increases as expected, but is still comparatively better than for the exact GP for $N=20$. All the other simulation study scenarios are presented as additional plots in the Section F (see Figures \ref{fig:m32-hyperparams}-\ref{fig:per-lowrho-hyperparams}).

\FloatBarrier
\section*{C: MCMC convergence diagnostics (additional figures)} \label{Supp:conv-diag}
We present the model convergence diagnostics of exact GPs and HSGPs for all the simulation scenarios except for the squared exponential (SE) scenario (which is shown in the main manuscript). In all of the scenarios, the HSGPs consistently the 1.01 threshold of model convergence as suggested by \cite{RankNorm_Vehtari_etal}. HSGPs subsequently also have much higher Bulk and Tail-ESS. Based on the diagnostics, HSGPs show much more consistent and stable convergence as compared to exact GPs.
\begin{figure}[!ht]
    \centering
    \includegraphics[width = \linewidth]{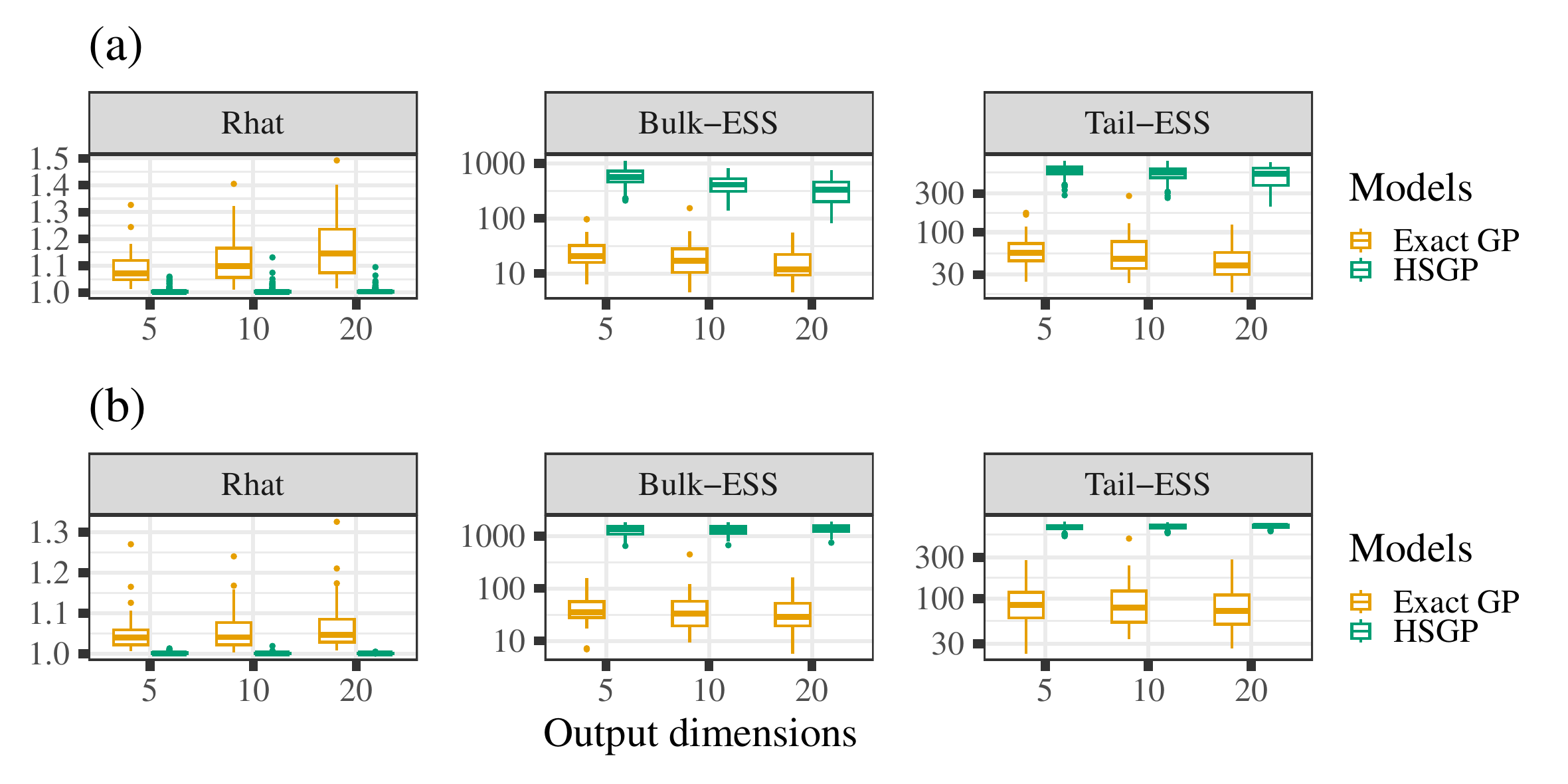}
    \caption{\textit{Matern 3/2 scenario: Convergence check for (a) latent inputs and (b) GP hyperparameters of the exact GPs (for $N = 20$) and HSGPs (combined for $N = 20, 50 \text{ } \text{and} \text{ } 200$ cases). The y-axes for Bulk and Tail ESS plots are log10 transformed.}}
    \label{fig:m32-exact-hsgp-valid}
\end{figure}

\begin{figure}[!ht]
    \centering
    \includegraphics[width = \linewidth]{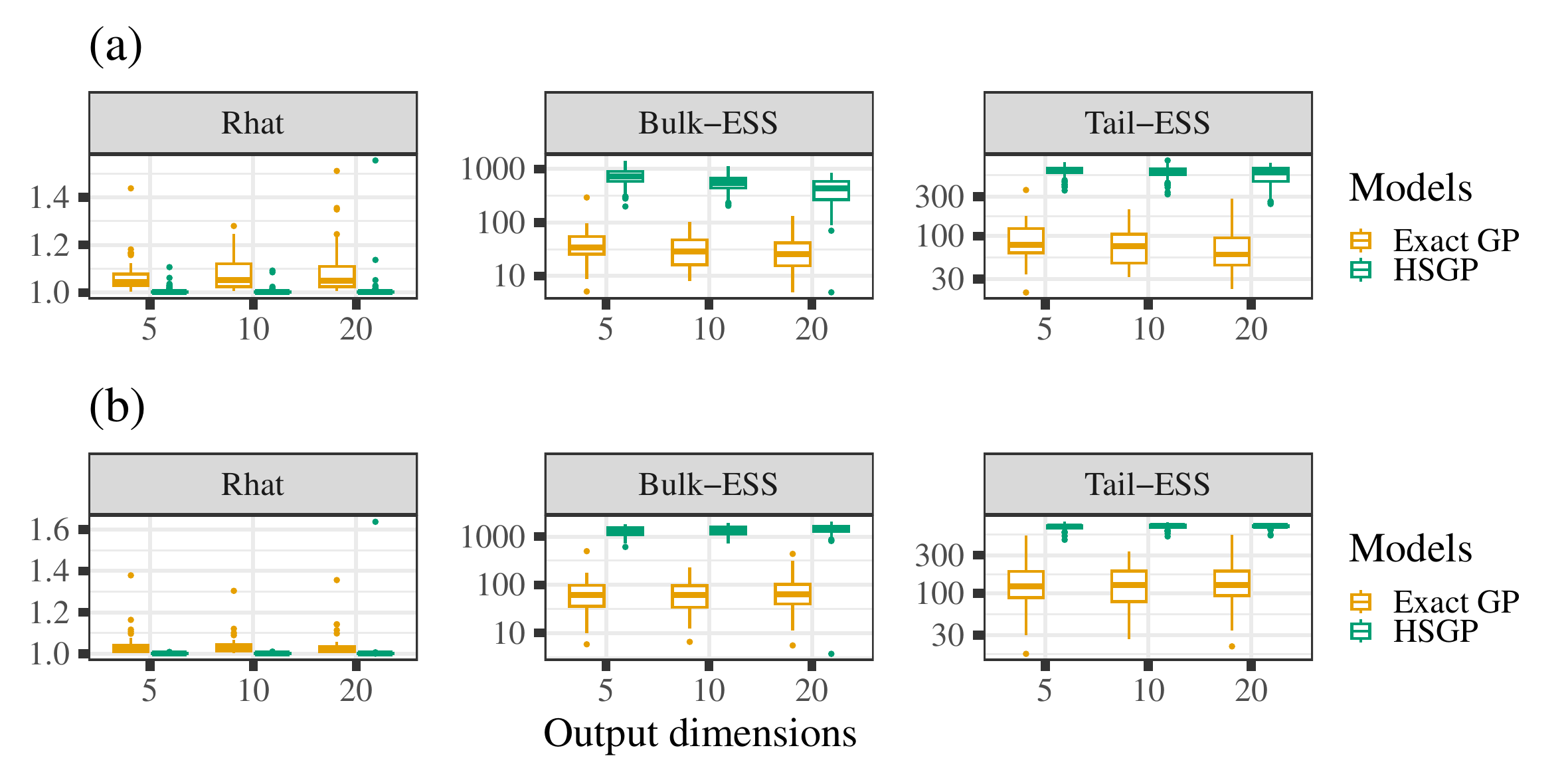}
    \caption{\textit{Matern 5/2 scenario: Convergence check for (a) latent inputs and (b) GP hyperparameters of the exact GPs (for $N = 20$) and HSGPs (combined for $N = 20, 50 \text{ } \text{and} \text{ } 200$ cases). The y-axes for Bulk and Tail ESS plots are log10 transformed.}}
    \label{fig:m52-exact-hsgp-valid}
\end{figure}

\begin{figure}[!ht]
    \centering
    \includegraphics[width = \linewidth]{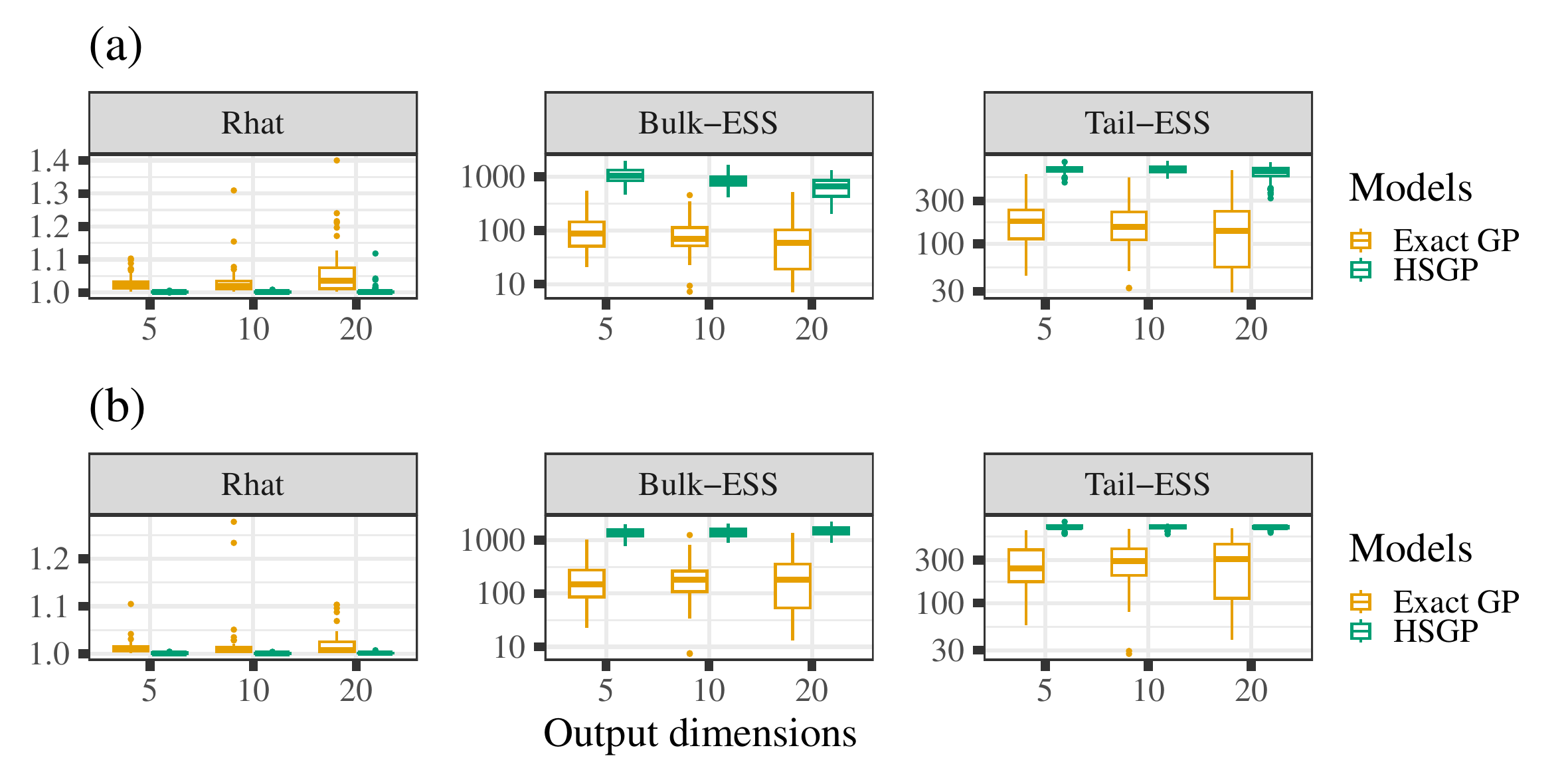}
    \caption{\textit{Periodic data scenario (lower oscillation): Convergence check for (a) latent inputs and (b) GP hyperparameters of the exact GPs (for $N = 20$) and HSGPs (combined for $N = 20, 50 \text{ } \text{and} \text{ } 200$ cases). The y-axes for Bulk and Tail ESS plots are log10 transformed.}}
    \label{fig:per-exact-hsgp-valid}
\end{figure}

\begin{figure}[!ht]
    \centering
    \includegraphics[width = \linewidth]{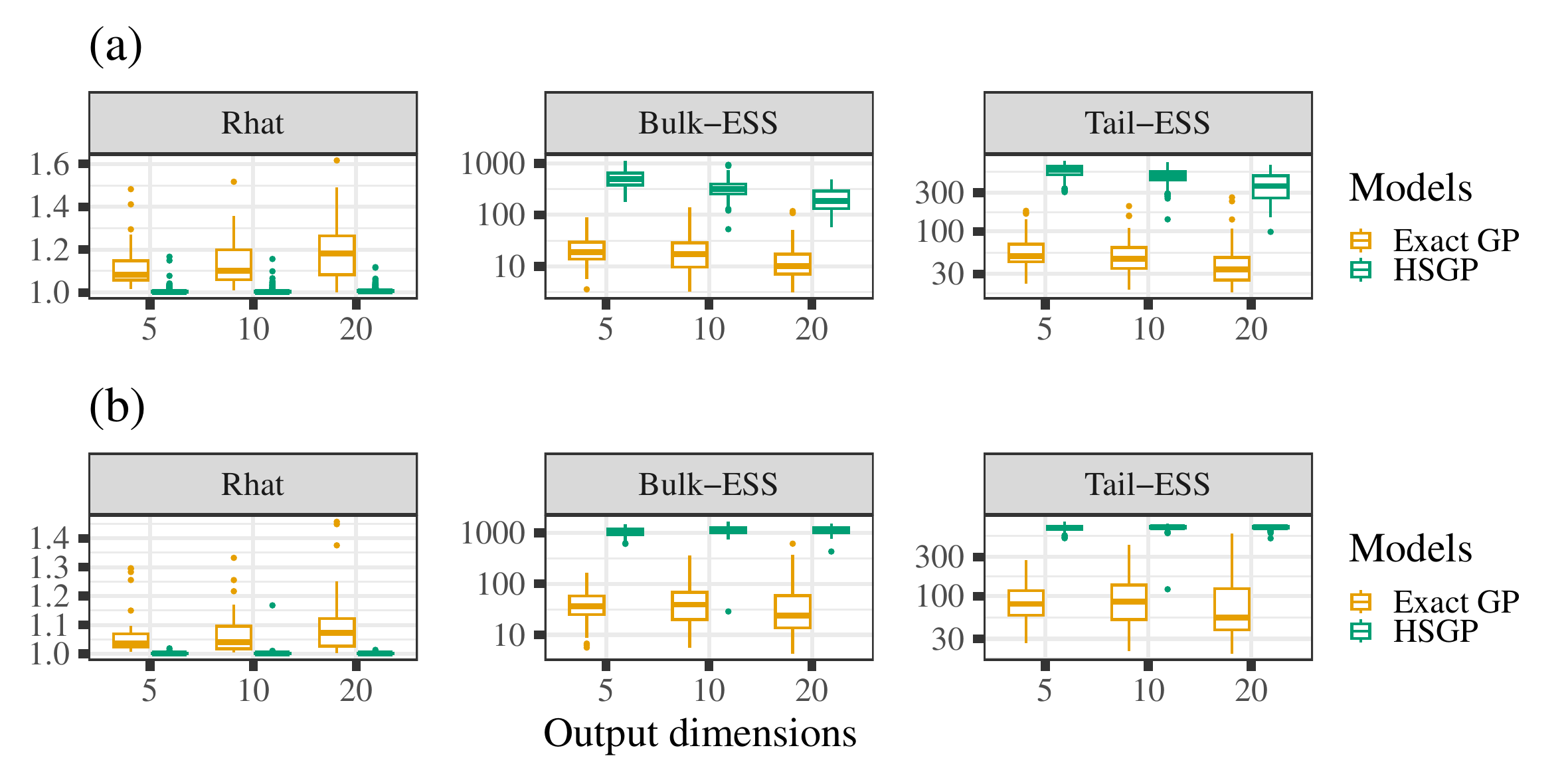}
    \caption{\textit{Periodic data scenario (higher oscillation): Convergence check for (a) latent inputs and (b) GP hyperparameters of the exact GPs (for $N = 20$) and HSGPs (combined for $N = 20, 50 \text{ } \text{and} \text{ } 200$ cases). The y-axes for Bulk and Tail ESS plots are log10 transformed.}}
    \label{fig:per-lowrho-exact-hsgp-valid}
\end{figure}

\begin{figure}[!ht]
    \centering
    \includegraphics[width = \linewidth]{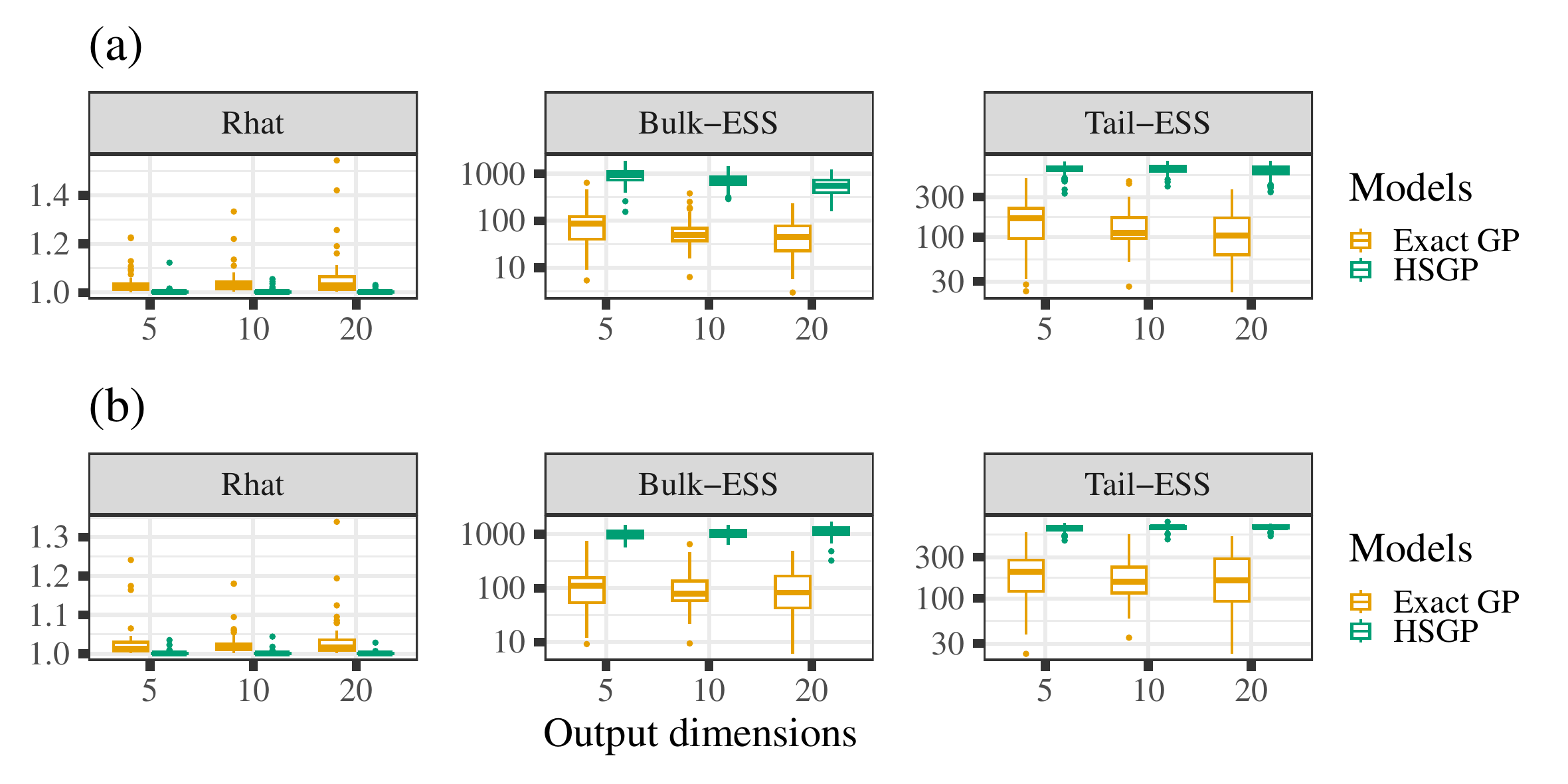}
    \caption{\textit{Squared exponential scenario (highly varying $\rho$): Convergence check for (a) latent inputs and (b) GP hyperparameters of the exact GPs (for $N = 20$) and HSGPs (combined for $N = 20, 50 \text{ } \text{and} \text{ } 200$ cases). The y-axes for Bulk and Tail ESS plots are log10 transformed.}}
    \label{fig:se-widerho-exact-hsgp-valid}
\end{figure}

\FloatBarrier
\section*{D: Further MCMC convergence analysis}
\begin{figure}[!ht]
    \centering
    \includegraphics[width = \linewidth]{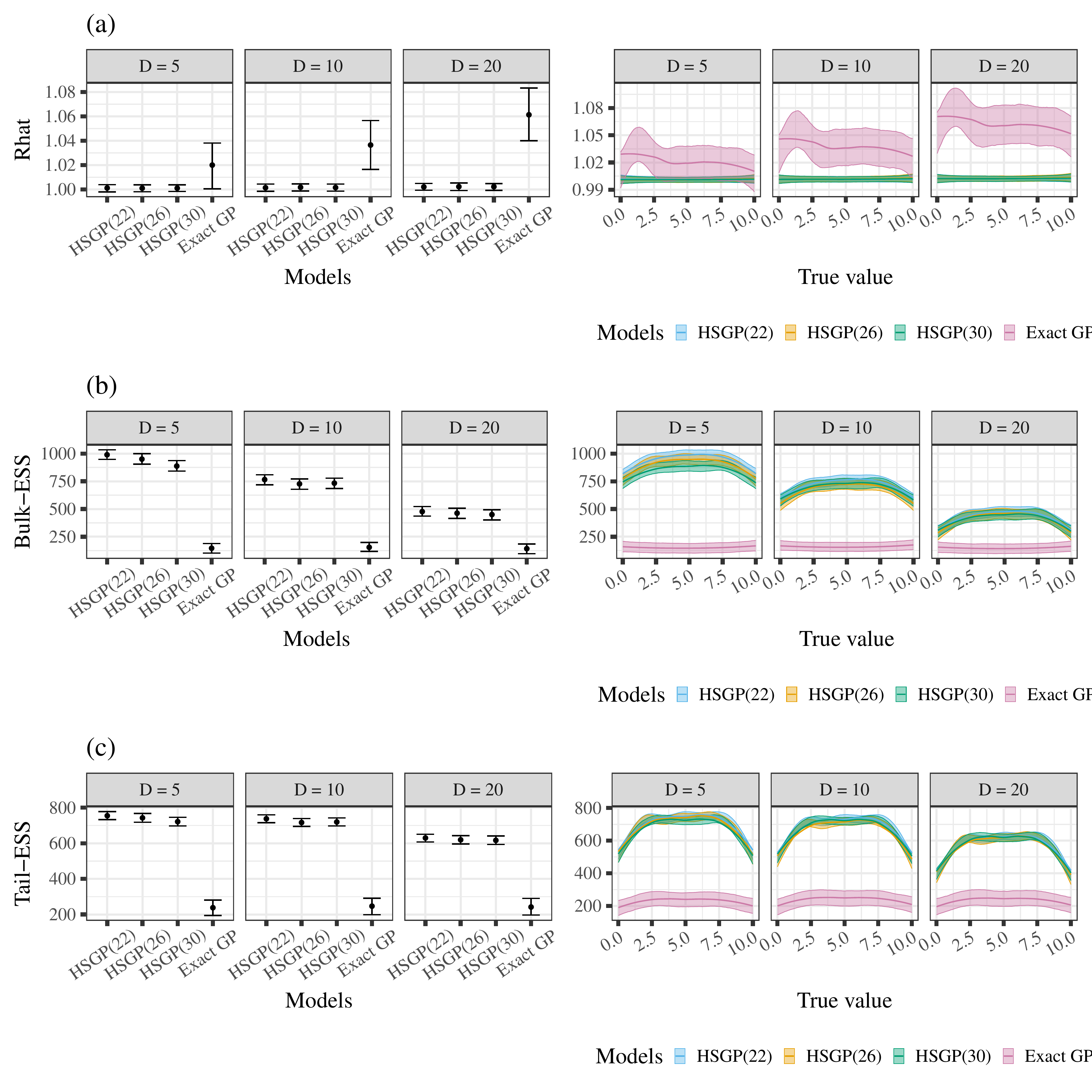}
    \caption{\textit{Squared exponential scenario: MCMC convergence comparison between exact GPs and HSGPs ($N = 20$ case) for latent inputs through (a) Rhat, (b) Bulk-ESS and (c) Tail-ESS measures. The behavior of each measures across ground truth are shown in the right-hand panel.}}
    \label{fig:se-convcheck-exact-hsgp-latentx}
\end{figure}

\begin{figure}[!ht]
    \centering
    \includegraphics[width = \linewidth]{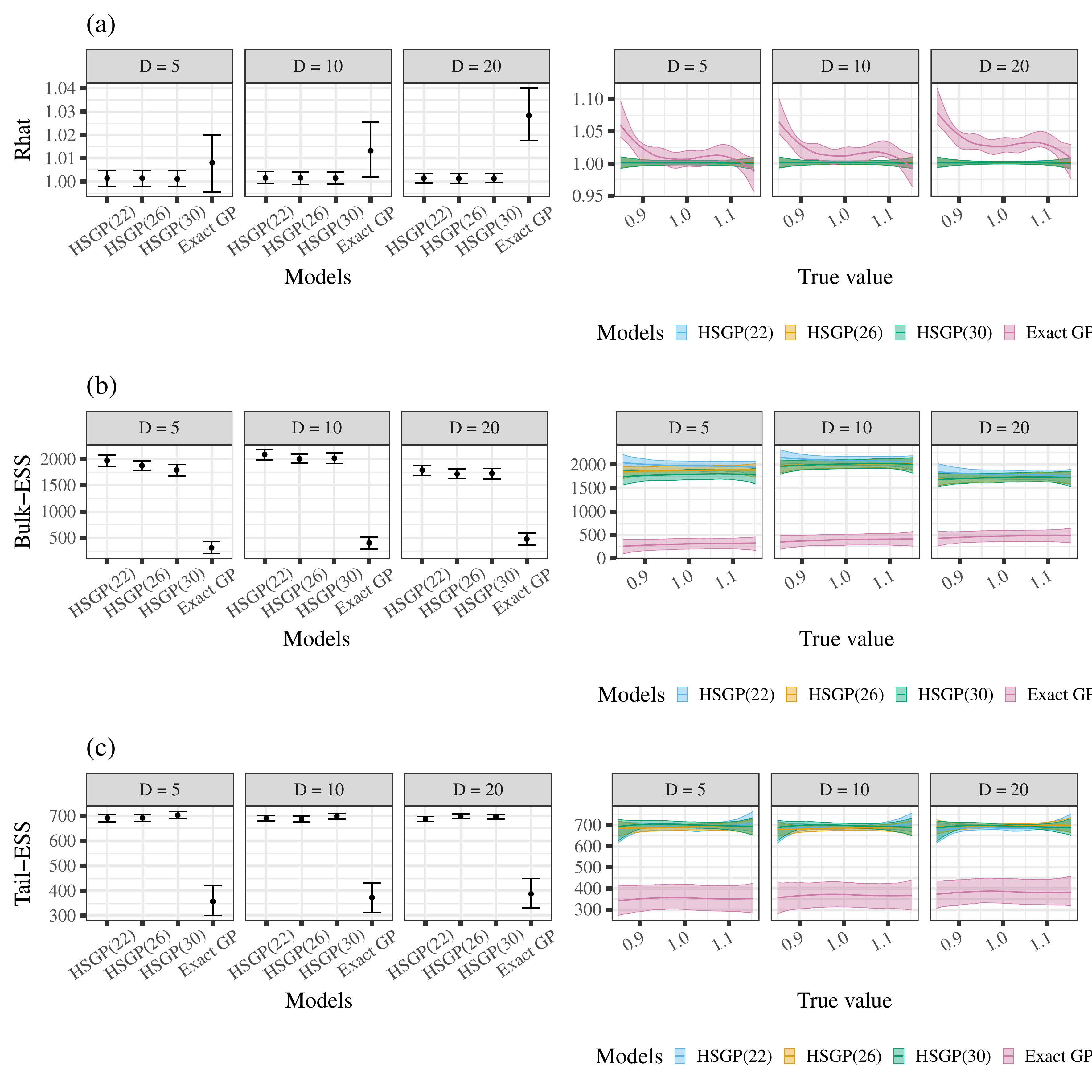}
    \caption{\textit{Squared exponential scenario: MCMC convergence comparison between exact GPs and HSGPs ($N = 20$ case) for length-scale through (a) Rhat, (b) Bulk-ESS and (c) Tail-ESS measures. The behavior of each measures across true values are shown in the right-hand panel.}}
    \label{fig:se-convcheck-exact-hsgp-rho}
\end{figure}

We further show MCMC convergence comparison for latent $x$, length-scale $\rho$, GP marginal SD $\alpha$ and error SD $\sigma$ between exact GPs and HSGPs for $N = 20$ case. Overall, HSGPs promise superior MCMC convergence compared to exact GPs, especially for latent variable inputs. Further, exact GPs struggle with convergence especially for smaller latent inputs as well as smaller length-scales as shown in the right panel of Figures \ref{fig:se-convcheck-exact-hsgp-latentx} and \ref{fig:se-convcheck-exact-hsgp-rho} respectively. As a consequence of these computational issues, exact GPs show miscalibrations in SBC checks and thus lower latent variable estimation accuracy compared to HSGPs.

\begin{figure}[!ht]
    \centering
    \includegraphics[width = \linewidth]{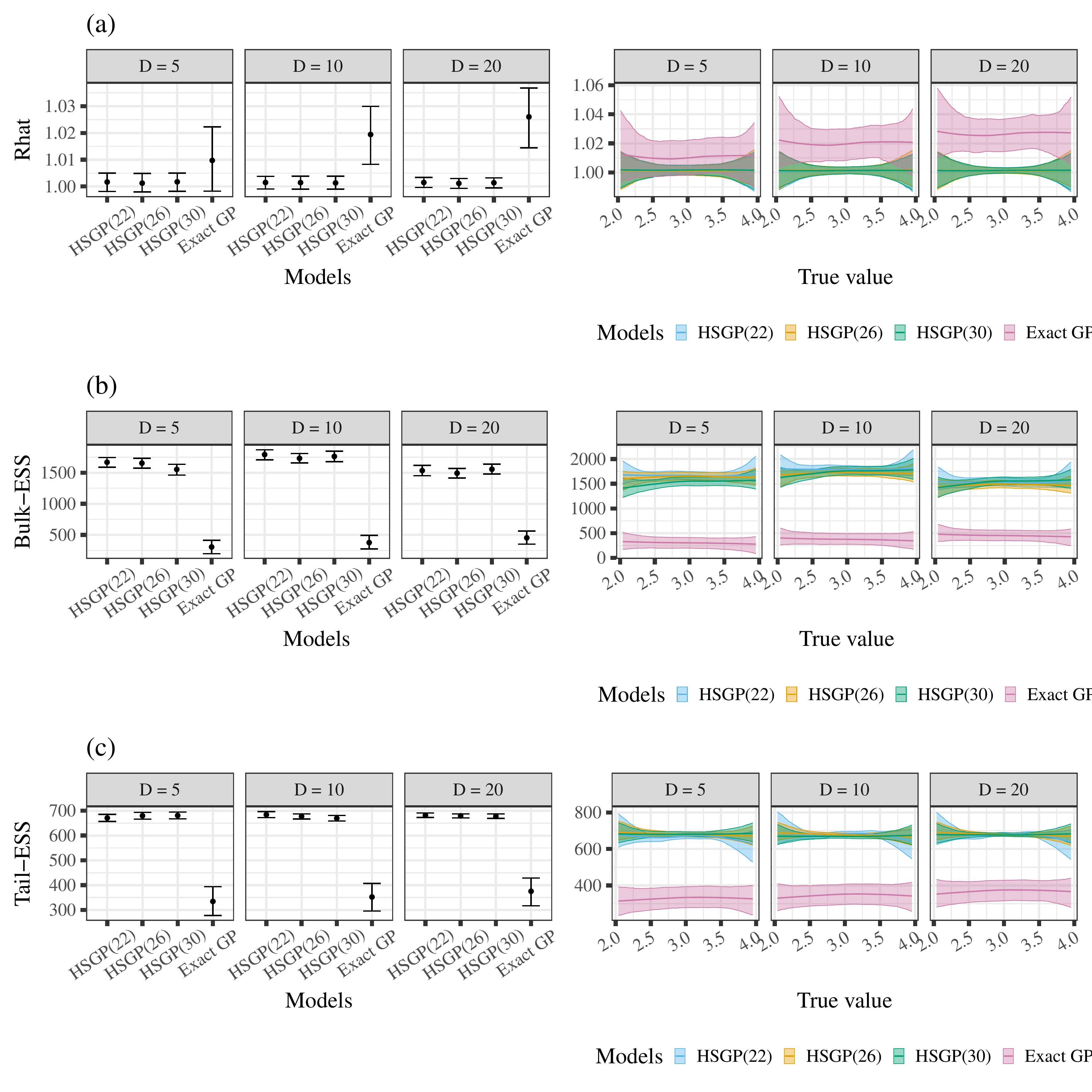}
    \caption{\textit{Squared exponential scenario: MCMC convergence comparison between exact GPs and HSGPs ($N = 20$ case) for GP marginal SD through (a) Rhat, (b) Bulk-ESS and (c) Tail-ESS measures. The behavior of each measures across ground truth are shown in the right-hand panel.}}
    \label{fig:se-convcheck-exact-hsgp-alpha}
\end{figure}

\begin{figure}[!ht]
    \centering
    \includegraphics[width = \linewidth]{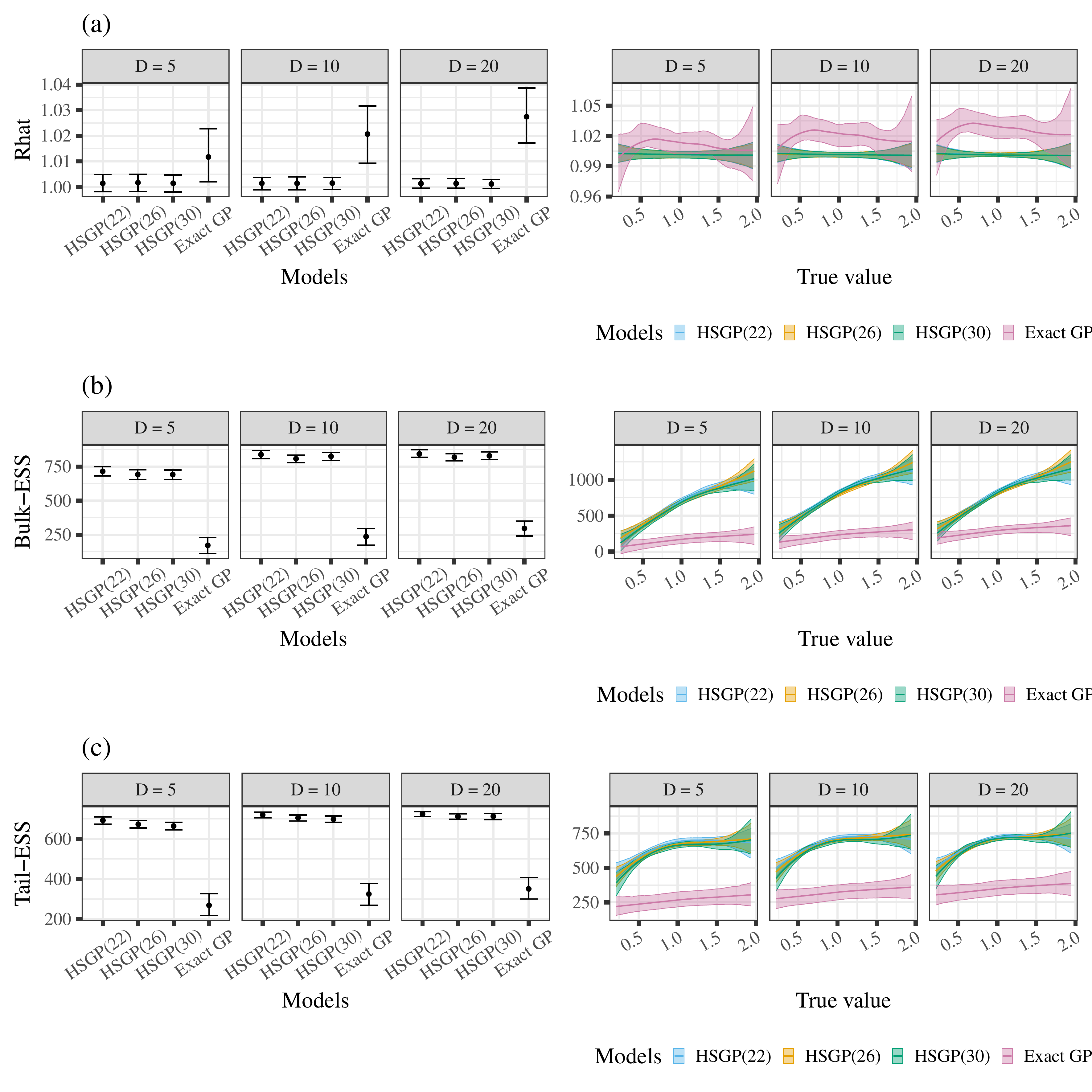}
    \caption{\textit{Squared exponential scenario: MCMC convergence comparison between exact GPs and HSGPs ($N = 20$ case) for error SD through (a) Rhat, (b) Bulk-ESS and (c) Tail-ESS measures. The behavior of each measures across ground truth are shown in the right-hand panel.}}
    \label{fig:se-convcheck-exact-hsgp-sigma}
\end{figure}

\FloatBarrier
\section*{E: Model calibration tests (additional figures)} \label{Supp:model-calib}
We present the ECDF plots for the fitted models in the Matern 3/2 and 5/2 simulation scenario with $N = 20$ and $D = 20$. Similar issues with calibration as in SE case presented in the main paper is seen with exact GPs and VIGPs.
\begin{figure}[!ht]
    \centering
    \includegraphics[width = \linewidth]{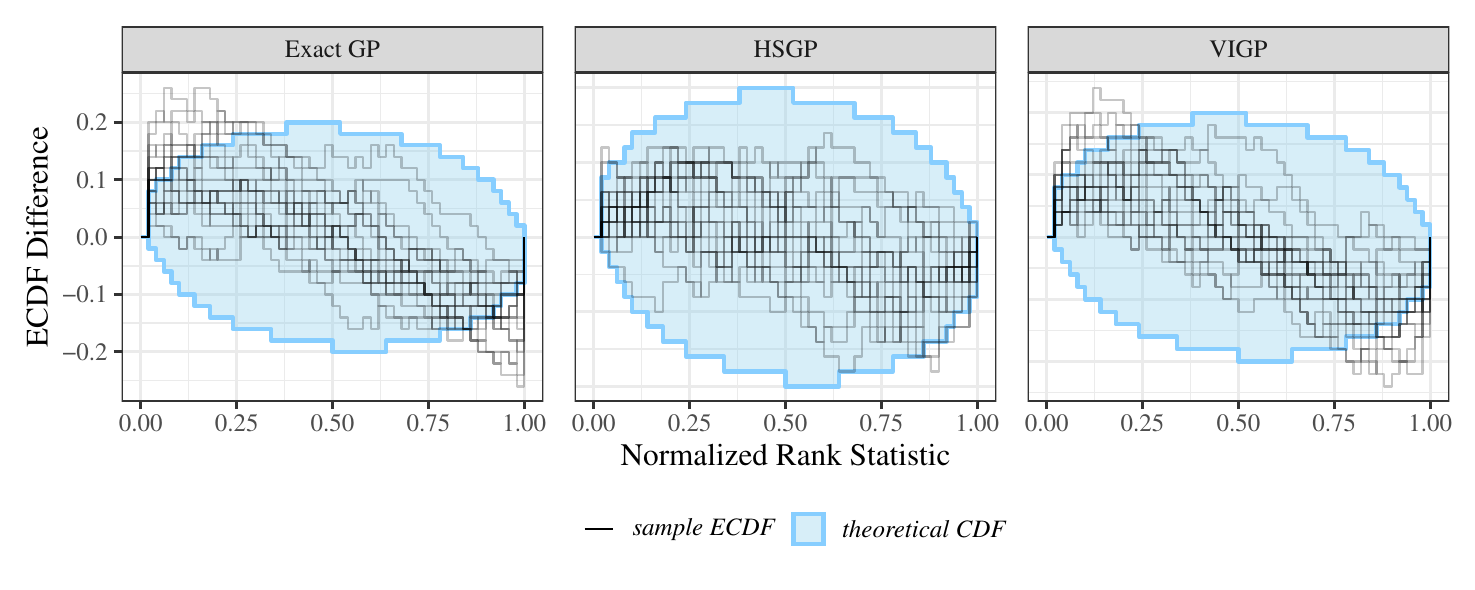}
    \caption{\textit{Squared exponential scenario: ECDF-difference calibration plots of the latent $x$ estimated by exact GP, HSGP, and VIGP. Only the HSGP is consistently well calibrated.}}
    \label{fig:m32-ecdf-n20-d20}
\end{figure}

\begin{figure}[!ht]
    \centering
    \includegraphics[width = \linewidth]{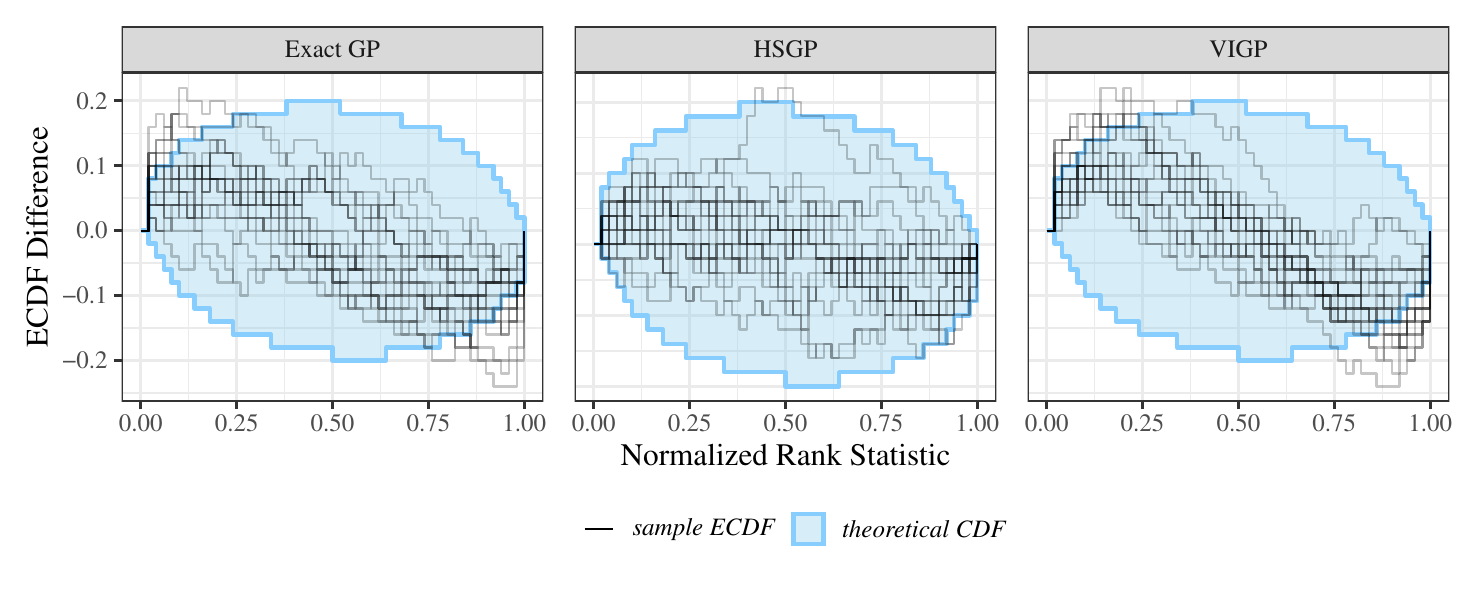}
    \caption{\textit{Squared exponential scenario: ECDF-difference calibration plots of the latent $x$ estimated by exact GP, HSGP, and VIGP. Only the HSGP is consistently well calibrated.}}
    \label{fig:m52-ecdf-n20-d20}
\end{figure}
We present the log $\gamma$ scores (offset by the 95\% confidence threshold) for all the fitted models in the Matern 3/2 and 5/2 simulation scenarios. The exact GPs and VIGPs consistently fail the calibration test for $D=20$. The HSGPs with $M=51$ number of representative basis functions show good calibration across all the scenarios. 
\begin{figure}[!ht]
    \centering
    \includegraphics[width = \linewidth]{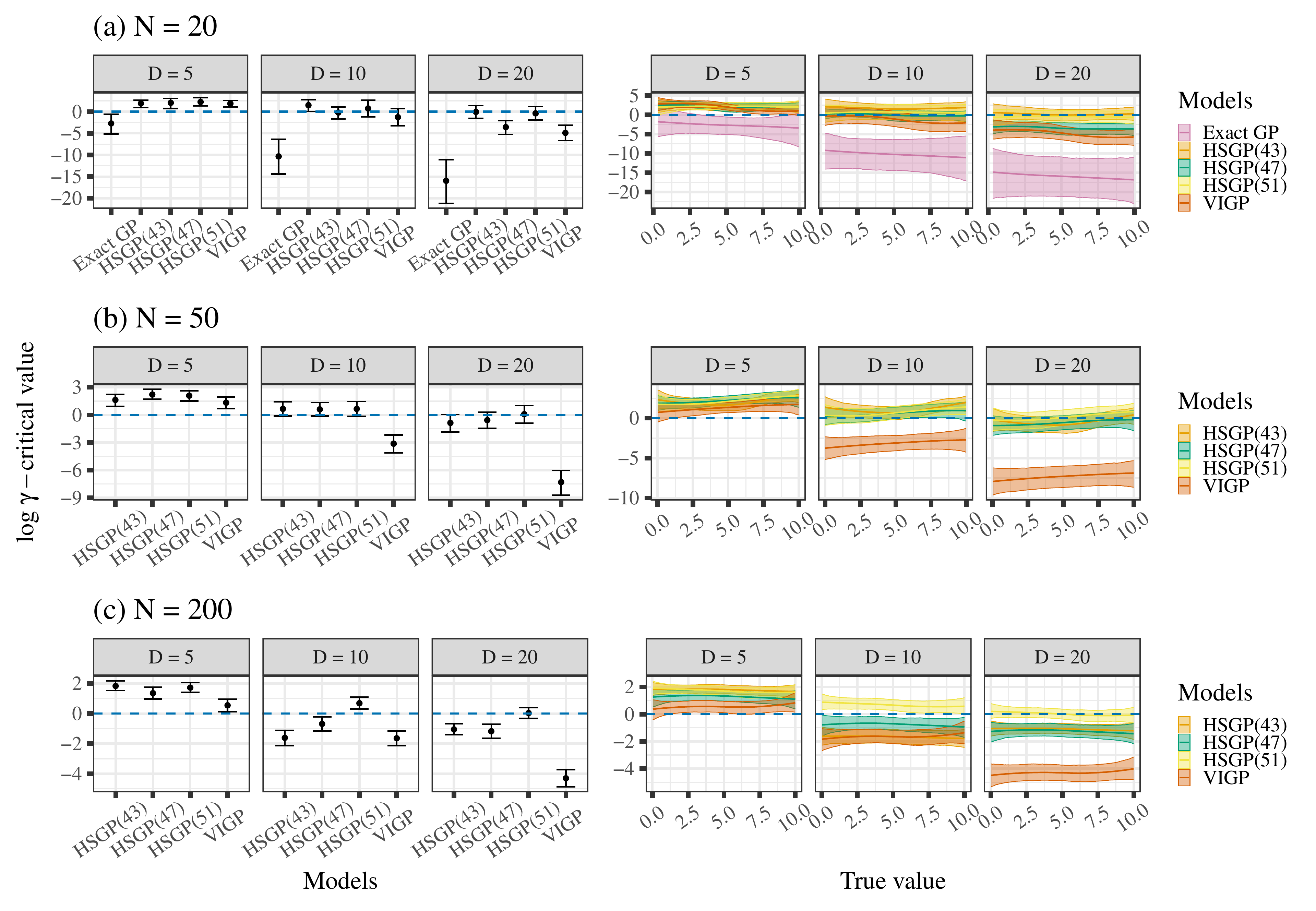}
    \caption{\textit{Matern 3/2 scenario: log $\gamma$ scores offset by the 95\% confidence threshold for all the fitted models. The behavior of scores across true latent $x$ values are shown in the right-hand panel. The blue dashed line denotes the threshold to reject uniformity, that is, models with values less than $0$ are miscalibrated. The HSGP($M$) shows the HSGPs with their corresponding number of basis functions.}}
    \label{fig:m32-log-gamma}
\end{figure}

\begin{figure}[!ht]
    \centering
    \includegraphics[width = \linewidth]{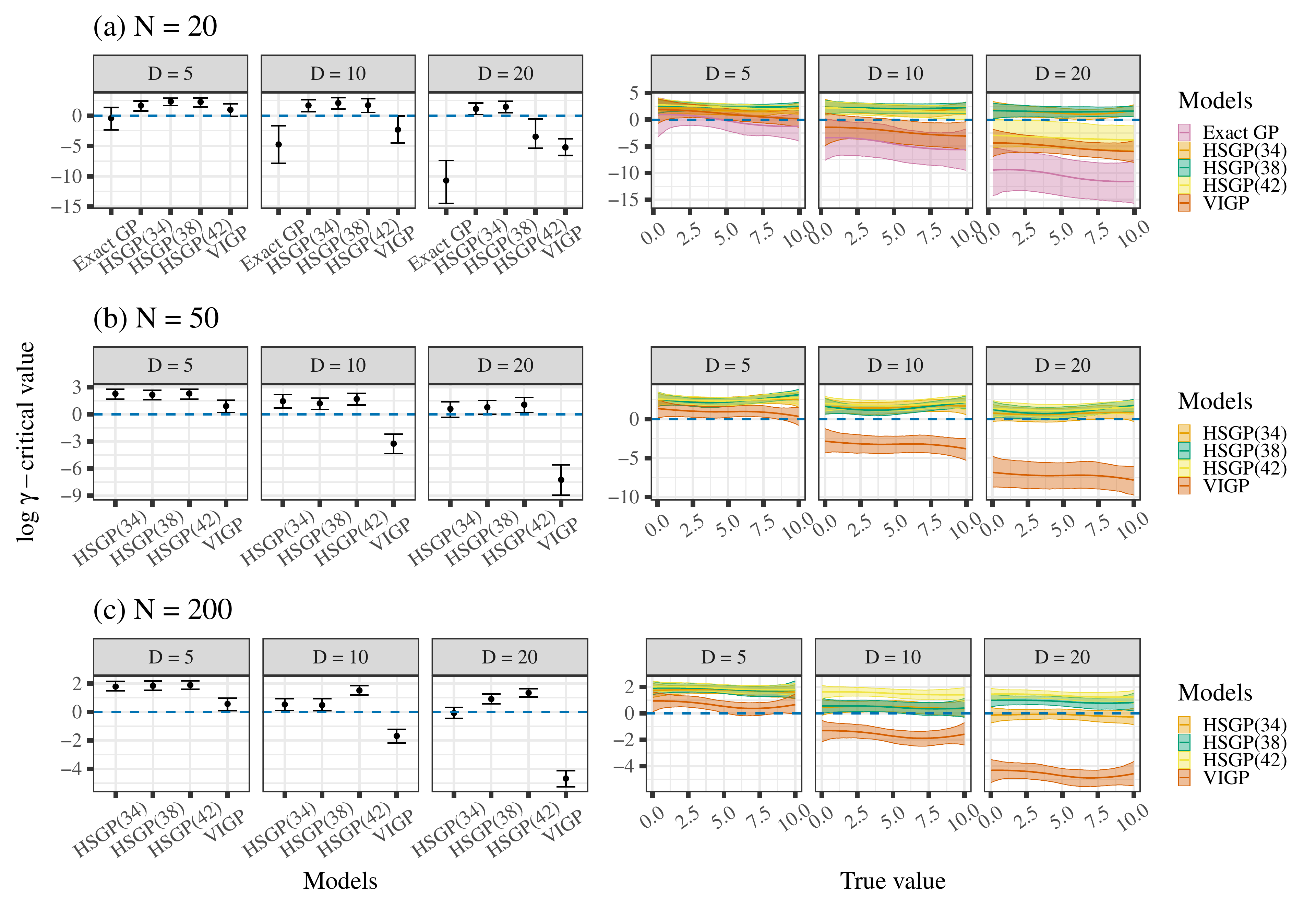}
    \caption{\textit{Matern 5/2 scenario: log $\gamma$ scores offset by the 95\% confidence threshold for all the fitted models. The behavior of scores across true latent $x$ values are shown in the right-hand panel. The blue dashed line denotes the threshold to reject uniformity, that is, models with values less than $0$ are miscalibrated. The HSGP($M$) shows the HSGPs with their corresponding number of basis functions.}}
    \label{fig:m52-log-gamma}
\end{figure}

\FloatBarrier
\section*{F: Latent input estimation (additional figures)} \label{Supp:latent-input-est}
We present the model evaluation on latent variable estimation accuracy for the Matern 3/2, 5/2 as well as the periodic data (both high and low oscillation) scenarios. In almost all of the cases, HSGPs have lower bias compared to exact GPs and VIGPs. In case of posterior SDs, the results are mixed. VIGPs sometimes show lower posterior SD as compared to HSGPs whereas in other cases, they are higher or same. Combined with the calibration test results, VIGPs are likely to underestimate posterior uncertainty due to severe model miscalibrations.
\begin{figure}[!ht]
    \centering
    \includegraphics[width = \linewidth]{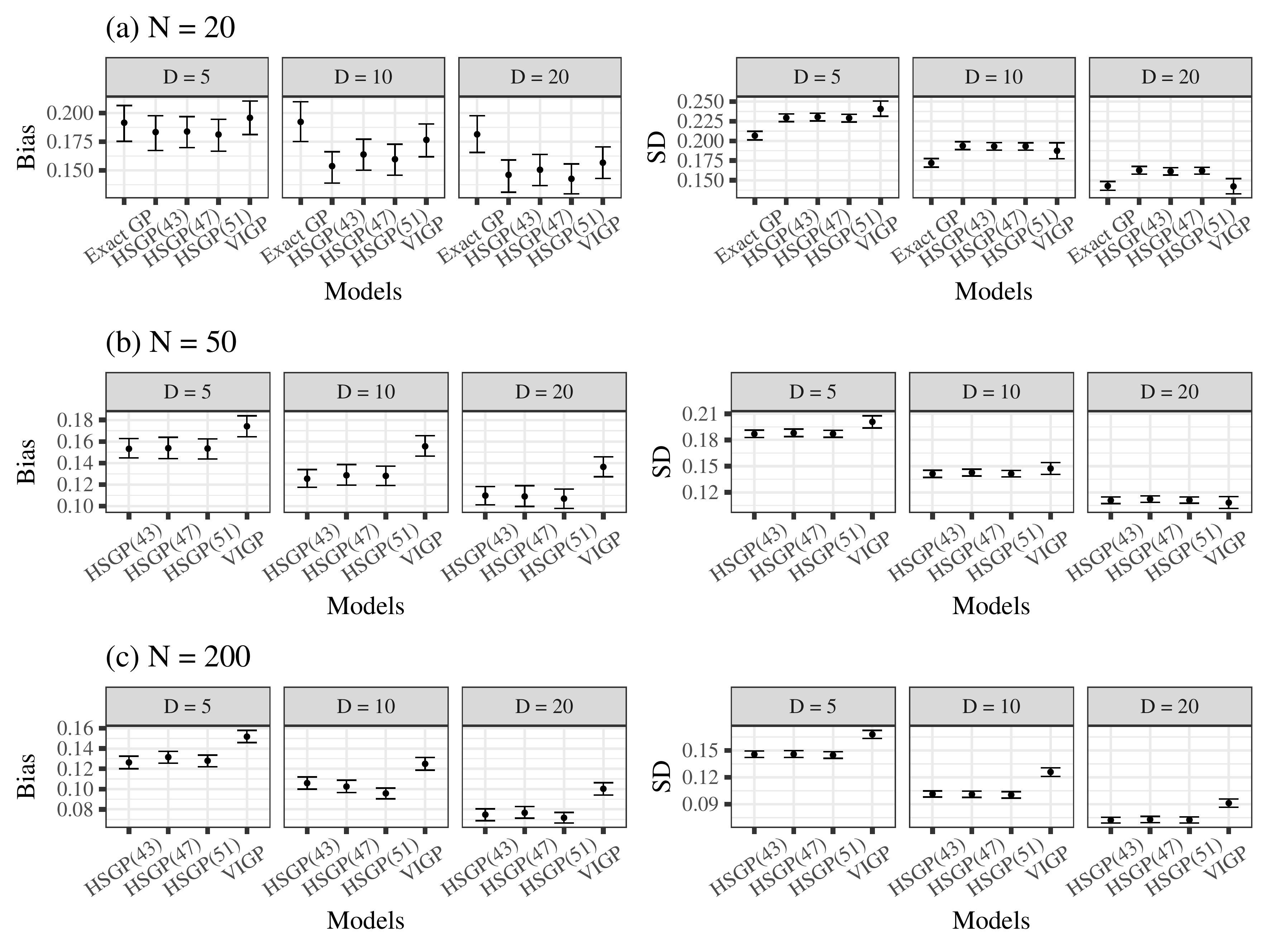}
    \caption{\textit{Matern 3/2 scenario: posterior bias and SD on recovery of latent inputs for all fitted models. The HSGP($M$) shows the HSGPs with their corresponding number of basis functions.}}
    \label{fig:m32-latentx}
\end{figure}

\begin{figure}[!ht]
    \centering
    \includegraphics[width = \linewidth]{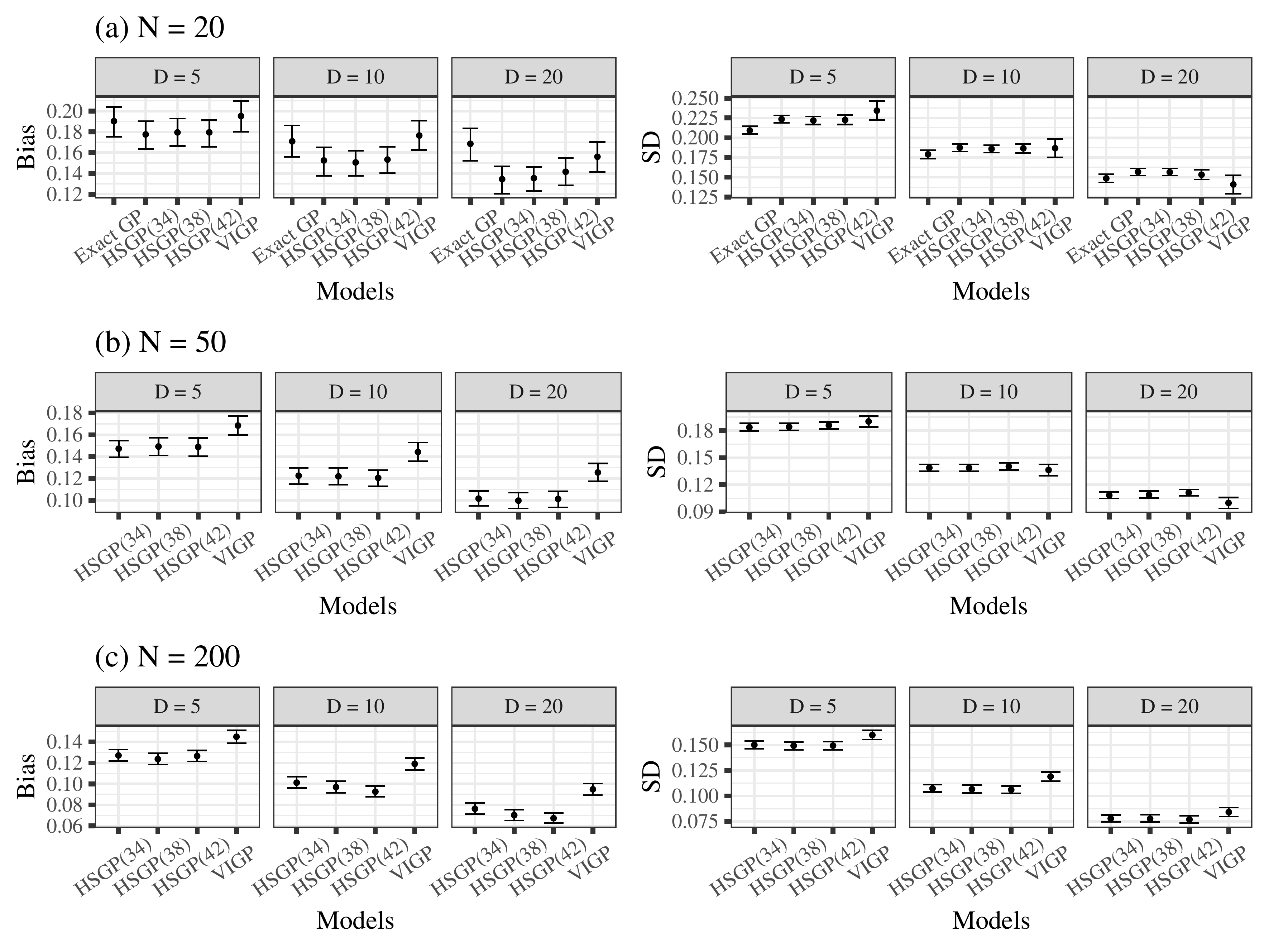}
    \caption{\textit{Matern 5/2 scenario: posterior bias and SD on recovery of latent inputs for all fitted models. The HSGP($M$) shows the HSGPs with their corresponding number of basis functions.}}
    \label{fig:m52-latentx}
\end{figure}

\begin{figure}[!ht]
    \centering
    \includegraphics[width = \linewidth]{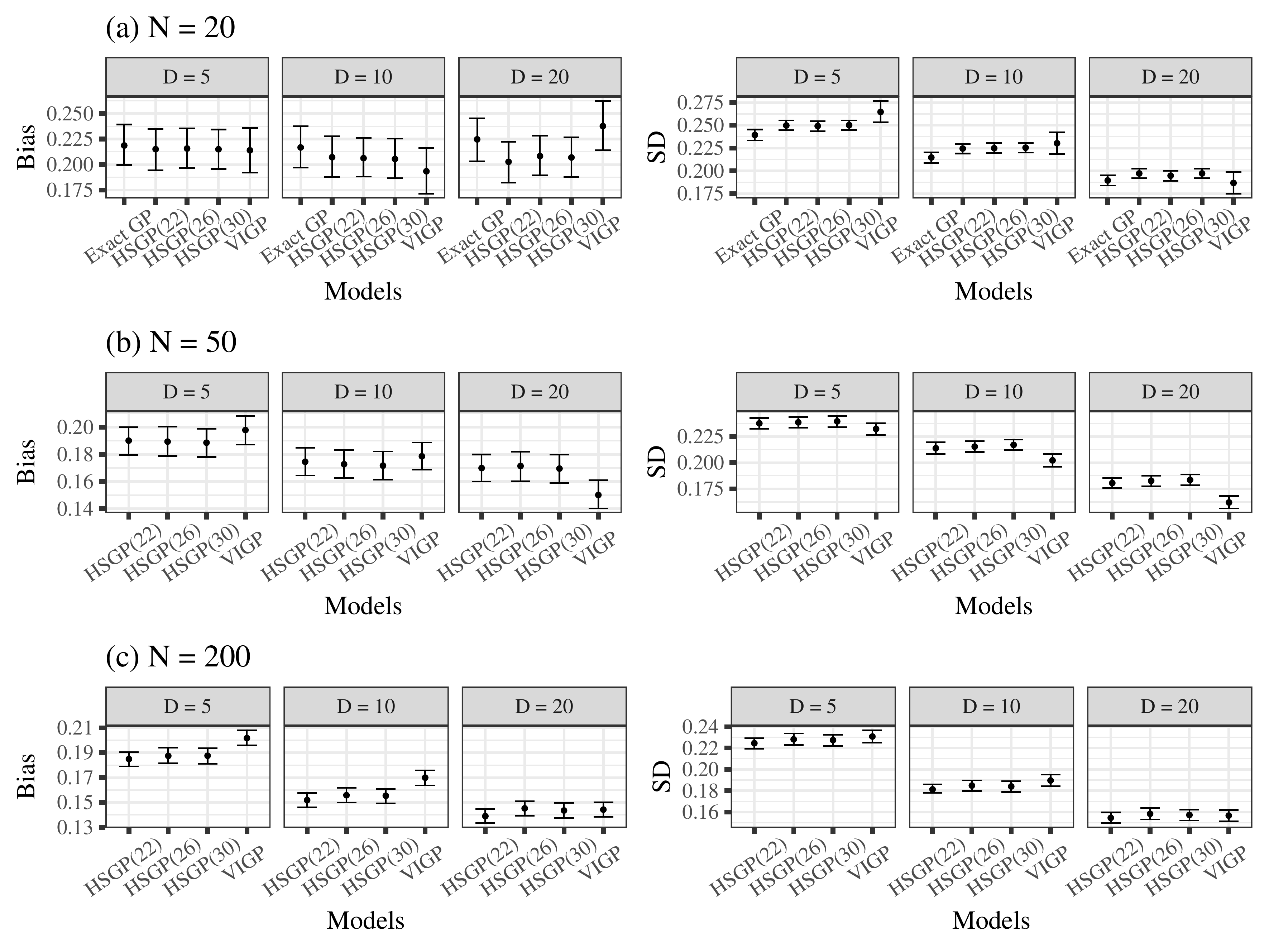}
    \caption{\textit{Periodic data scenario (lower oscillations): posterior bias and SD on recovery of latent inputs for all fitted models. The HSGP($M$) shows the HSGPs with their corresponding number of basis functions.}}
    \label{fig:per-latentx}
\end{figure}

\begin{figure}[!ht]
    \centering
    \includegraphics[width = \linewidth]{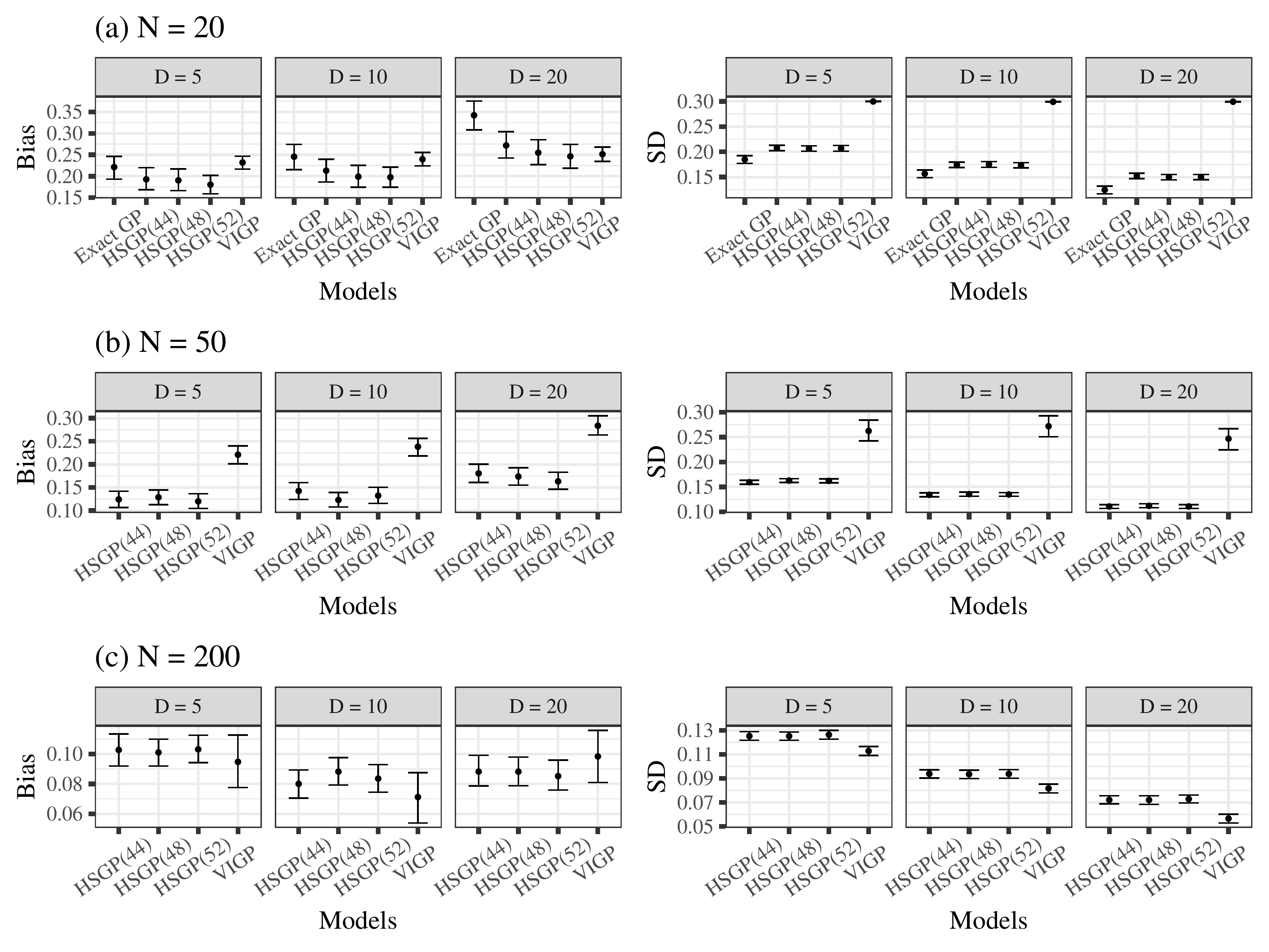}
    \caption{\textit{Periodic data scenario (higher oscillations): posterior bias and SD on recovery of latent inputs for all fitted models. The HSGP($M$) shows the HSGPs with their corresponding number of basis functions.}}
    \label{fig:per-lowrho-latentx}
\end{figure}

\FloatBarrier
\section*{G: Checking robustness of HSGPs with wide priors (additional figures)}
\begin{figure}[!ht]
    \centering
    \includegraphics[width = \linewidth]{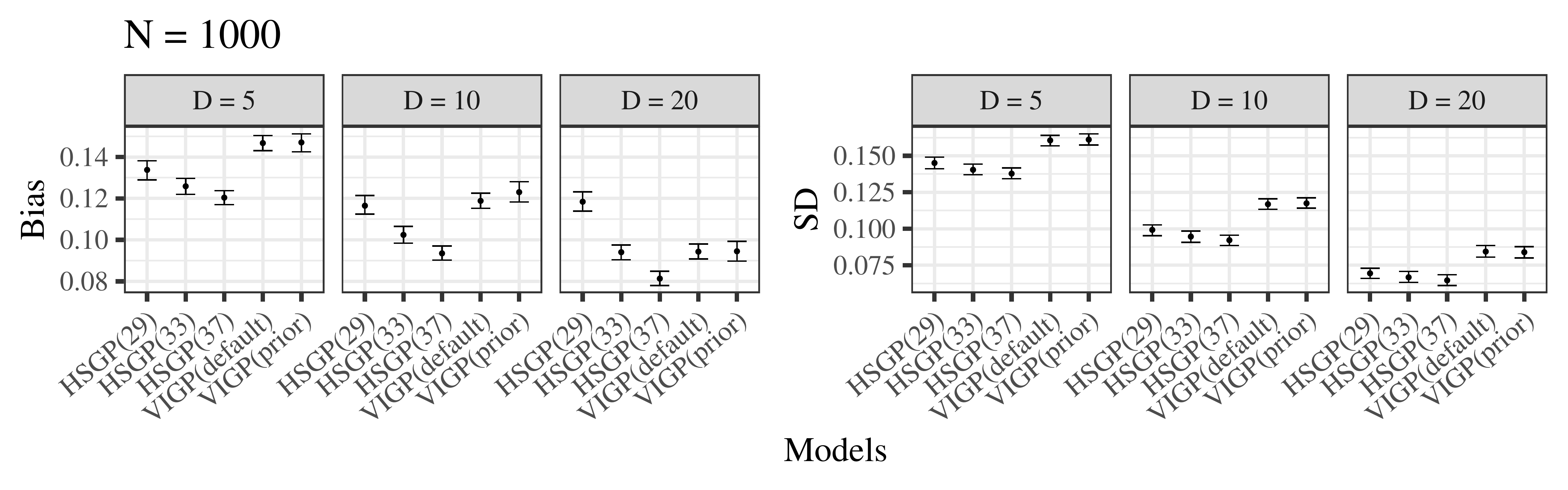}
    \caption{\textit{Matern 3/2 scenario (special case): posterior bias and SD on recovery of latent inputs for all fitted models. The HSGP($M$) shows the HSGPs with their corresponding number of basis functions. VIGP (default) shows the default Pyro implementation and VIGP (prior) has fixed GP hyperparameters informed by the true sampling distributions.}}
    \label{fig:m32-n1000-latentx}
\end{figure}

\begin{figure}[!ht]
    \centering
    \includegraphics[width = \linewidth]{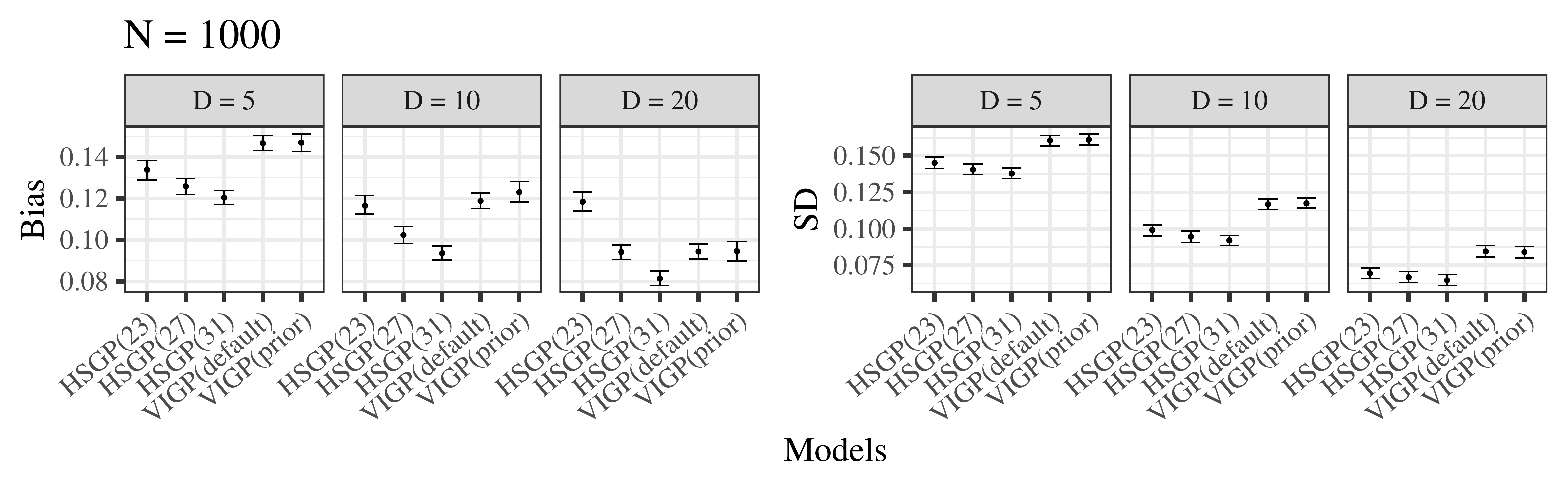}
    \caption{\textit{Matern 5/2 scenario (special case): posterior bias and SD on recovery of latent inputs for all fitted models. The HSGP($M$) shows the HSGPs with their corresponding number of basis functions. VIGP (default) shows the default Pyro implementation and VIGP (prior) has fixed GP hyperparameters informed by the true sampling distributions.}}
    \label{fig:m52-n1000-latentx}
\end{figure}

\FloatBarrier
\section*{H: Hyperparameter estimation (additional figures)} \label{Supp:add-hyperparams-est}
The GP hyperparameter estimation results for the Matern 3/2, 5/2  and the periodic data scenarios (both lower and higher oscillations) are presented here. The results are qualitatively similar across all the simulation scenarios.

\section*{I: List of genes for the case study}
In our real-world case study for cell cycle data, we use the data from all the cells along with a selection of 12 influential genes based on the recommendations of \cite{mahdessian_spatiotemporal_2021}. The list of genes used are: CCNA2, CCNB1, BIRC5, TOP2A, PLK1, NDC80, CDCA8, MKI67, CDC6, CENPF, KRT17, RRM2. These recommendations were primarily concerned with high amount of RNA activities denoted by their counts.
\FloatBarrier

\end{document}